\documentclass[final,3p,times]{elsarticle}

\usepackage[utf8]{inputenc}
\usepackage{amsmath,amssymb,amsthm,amsfonts,eucal,dsfont,mathrsfs,latexsym}
\usepackage{graphicx} 
\usepackage{booktabs}
\usepackage{caption}
\usepackage{subcaption}
\usepackage{hyperref}
\usepackage{calrsfs}

\hypersetup{
    bookmarks=true,         
    unicode=false,          
    pdftoolbar=true,        
    pdfmenubar=true,        
    pdffitwindow=false,     
    pdfstartview={FitH},    
    pdftitle={A Generalized Convolution Model and Estimation for Non-stationary Random Functions},
    pdfauthor={Francky Fouedjio},     
    pdfsubject={Non-stationary Random Functions},   
    pdfcreator={Francky Fouedjio},   
    pdfproducer={Francky Fouedjio}, 
    pdfkeywords={non-stationarity} {convolution} {covariance} {kernel} {kriging} {simulation} {covariance}, 
    pagebackref=true,
    pdfnewwindow=true,      
    colorlinks=true,       
    linkcolor=red,          
    citecolor=green,        
    filecolor=magenta,      
    urlcolor=cyan           
}

\usepackage{sectsty}
\usepackage[compact]{titlesec} 
\allsectionsfont{\scshape\selectfont}

\small\normalsize
\newtheorem{theorem}{Theorem}[section]
\newtheorem{lemma}[theorem]{Lemma}
\newtheorem{prop}[theorem]{Proposition}
\newtheorem{cor}[theorem]{Corollary}

\theoremstyle{definition}

\newtheorem{example}[theorem]{Example}
\theoremstyle{remark}

\DeclareMathOperator*{\argmin}{arg\,min}

\journal{ }

\begin{document}

\begin{frontmatter}

\title{A Generalized Convolution Model and Estimation for Non-stationary Random Functions}

\author[]{F. Fouedjio}
\ead{francky.fouedjio@mines-paristech.fr}

\author[]{N. Desassis}
\ead{nicolas.desassis@mines-paristech.fr}

\author[]{J. Rivoirard}
\ead{jacques.rivoirard@mines-paristech.fr}

\address{\'{E}quipe G\'{e}ostatistique - Centre de G\'{e}osciences, MINES ParisTech}
\address{35, rue Saint-Honor\'{e}, 77305 Fontainebleau, France}

\begin{abstract}
Standard geostatistical models assume second order stationarity of the underlying Random Function. In some instances, there is little reason to expect the spatial dependence structure to be stationary over the whole region of interest. In this paper, we introduce a new model for second order non-stationary Random Functions as a convolution of an orthogonal random measure with a spatially varying random weighting function. This new model is a generalization of the common convolution model where a non-random weighting function is used. The resulting class of non-stationary covariance functions is very general, flexible and allows to retrieve classes of closed-form non-stationary covariance functions known from the literature, for a suitable choices of the random weighting functions family. Under the framework of a single realization and local stationarity, we develop parameter inference procedure of these explicit classes of non-stationary covariance functions. From a local variogram non-parametric kernel estimator, a weighted local  least-squares approach in combination with kernel smoothing method is developed to estimate the parameters. Performances are assessed on two real datasets: soil and rainfall data. It is shown in particular that the proposed approach outperforms the stationary one, according to several criteria.  Beyond the spatial predictions, we also show how conditional simulations can be carried out in this non-stationary framework.
\end{abstract}

\begin{keyword}
non-stationarity \sep convolution \sep covariance \sep kernel \sep kriging \sep simulation.
\end{keyword}

\end{frontmatter}

\section{Introduction}
\label{sec1}

Modelling and estimating the underlying spatial dependence structure of the observed data is a key element to spatially interpolate data and perform simulations. Its description is commonly carried out using statistical tools such as the covariance or variogram calculated on the whole domain of interest. Simplifying assumptions are often made on the spatial dependence structure. They include  stationarity assumption where the second order association between pairs of locations is assumed to depend only on the vector between these locations. However, it has become increasingly clear that this assumption is driven more by mathematical convenience than by reality.  In pratice, it often happens that the stationarity assumption may be doubtful. Non-stationarity can occur due to many factors, including specific landscape and topographic features of the region of interest or other localized effects. These local influences can be observed by computing local variograms, whose characteristics may vary across the domain of observations. More practionners choose to replace this convenient assumption with more realistic assumption of non-stationarity. They subdivide the modelling area into stationary domains, thereby increasing the required professional time for modelling and potentially producing disjointed domains that are globally inconsistent. 

Various approaches have been developped over the years to deal with non-stationarity through second order moments (see \citep{Gut13,Sam01,Gut94}, for a review ). One of the most popular methods of introducing non-stationarity is the convolution approach. It consists of taking a spatial white noise, then averaging it using weights that vary spatially to thereby obtain a non-stationary Random Function. In this way, the resulting spatial dependence structure is non-stationary. The covariance or variogram as a whole is allowed to vary in some way between different locations. \citet{Hig98b} use spatially varying Gaussian kernel function to induce a non-stationary covariance structure. The resulting covariance function has a closed-form, but is infinitely differentiable, which may not be desirable for modelling real phenomena. \citet{Zhu10} use a family of spatially varying modified Bessel kernel functions to produce a non-stationary covariance function with local smoothness characteristics that are similar to the stationary Matérn covariance functions class. One limitation of this approach is that the non-stationary covariance function does not have a closed-form and in general can only be evaluated by numerical integration. Moreover, this approach does not take into account a spatially varying anisotropy. 

Furthermore, explicit expressions of some non-stationary covariance functions class that include a non-stationary version of the stationary Matérn covariance function have been introduced by  \citet{Pac06}. This was further developed by \citet{Ste05}, \citet{Por09} and \citet{Mat10}. However, these classes of analytical non-stationary covariance functions do not directly derived from a Random Function model like the convolution representation or the spectral representation. Thus, this does not facilitate their understanding and interpretability. It is helpful to have a constructive approach for Random Functions admitting such closed-form non-stationary covariance functions. Moreover, the inference of the parameters of these latter remains a critical problem. \citet{Pac06} enumerate some difficulties and suggest some possible methods including the moving windows approach based on the local variogram or local likelihood. \citet{And11} mention two typical problems arising with moving windows method. First, the range of validity of a stationary approximation can be too small to contain enough local data to estimate the spatial dependence structure reliably. Second, it can produce non-smooth parameter estimates, leading to discontinuities on the kriging map which is undesirable in many cases. \citet{And11} develop a weighted local likelihood approach for estimating parameters of the non-stationary version of the stationary Matérn covariance function. This approach downweight the influence of distant observations and allows irregular sampling observations. A few drawbacks of their approach are the computational burden of inverting covariance matrices at every location for parameter estimation and the Gaussian distributional assumption  for analytical tractability.

In this work, we are interested primarily in the modelling of second order non-stationary Random Functions, using a constructive approach in the spirit of convolution model. We present a new model for second order non-stationary Random Functions  as a convolution of an orthogonal random measure, with a spatially varying stochastic weighting function. This is an extension of the classical convolution model where a deterministic weighting function is used. By this modelling approach, the resulting class of non-stationary covariance functions is very general, flexible and allows to retrieve classes of closed-form non-stationary spatial covariance functions known from the literature, for a suitable choices of the stochastic weighting functions family. These latter classes show locally a stationary behaviour and their parameters are allowed to vary with location, yielding local variance, range, geometric anisotropy and smoothness. In this way, it naturally allows stationary local covariance parameters to be knitted together into a valid non-stationary global covariance. Thus, we establish a direct link between the existing explicit classes of non-stationary covariance functions and the convolution approach. This construction bears some resemblance to the moving average model with stochastic weighting function introduced by \citet{Mat86} to generate stationary covariance functions like Cauchy and Matérn families. Secondly, we develop a procedure of estimating parameters that govern these classes of closed-form non-stationary covariance functions  under a single realization and local stationarity  framework, through a step by step approach. First, we compute local variograms by a non-parametric kernel estimator.  Then, it is used in a weighted local least squares procedure for estimating the parameters at a reduced set of representative points referred to as anchor points. Finally, a kernel smoothing method is used to interpolate the parameters at any location of interest.  The estimation method proposed is free distribution and no matrix inversions or determinants calculation are required. It is  applied to two real datasets: soil and rainfall data.

The outline  of the paper is as follows: a  generalized convolution model is described through its basic ingredients and main properties in Section \ref{sec2}. In Section \ref{sec3}, we show how the model allows us to construct explicit classes of non-stationary covariance functions.  In Section \ref{sec4}, parameter inference of these explicit classes of non-stationary covariance is detailed. We tackle the spatial predictions and conditional simulations in Section \ref{sec5}. In Section \ref{sec6},  two real datasets are used to illustrate the performances of the proposed method and its potential. Finally, Section \ref{sec7} outlines concluding remarks. 

\section{Model Definition}
\label{sec2}

Let $Z=\{Z(\mathbf{x}): \mathbf{x} \in G \subseteq \mathds{R}^p, p\geq 1\}$  be a  Random Function  defined on a fixed continuous domain of interest $G$ of the Euclidean space $\mathds{R}^p$. The Random Function $Z$ is defined as follows:
\begin{equation}{\label{Eq1}}
Z(\mathbf{x})=\int_{{\mathds{R}}^p}f_{\mathbf{x}}(\mathbf{u};T(\mathbf{u}))W(d\mathbf{u}),  \quad  \forall \mathbf{x} \in G,
\end{equation}
where $W(.)$ is an orthogonal random measure on $\mathds{R}^p$ with $\mathds{E}(W(d\mathbf{u}))=0, \  \mathds{E}\left( W(d\mathbf{u}) W(d\mathbf{v})\right)=\lambda \delta_{\mathbf{u}}(d\mathbf{v})d\mathbf{u}, \ \forall\mathbf{u}, \mathbf{v} \in \mathds{R}^p$; $\delta_{\mathbf{u}}(.)$ is the Dirac measure at $\mathbf{u}$ and $\lambda \in \mathds{R}_+^*$. $\{T(\mathbf{u}), \mathbf{u} \in \mathds{R}^p\}$ is a family of independent and identically distributed random variables, and independent of $W(.)$,  taking their values in the subset $\mathcal{T}$ of $\mathds{R}$ according to a probability measure $\mu(.)$.
$\{\mathbf{u} \mapsto f_{\mathbf{x}}(\mathbf{u};t), \mathbf{x} \in G \subseteq \mathds{R}^p\}$ is a family of square-integrable functions on $\mathds{R}^p$, for all $t \in \mathcal{T}$. For fixed $ \mathbf{u}$,  $f_{\mathbf{x}}(\mathbf{u};T(\mathbf{u}))$ is a random variable whose first two  moments are assumed finite and integrable on $\mathds{R}^p$.

Under these assumptions, the stochastic convolution defined in (\ref{Eq1}) is well-defined in $L^2$, that is $\mathds{V}(Z(\mathbf{x}))=\lambda\int_{\mathds{R}^p}\mathds{E}\{f_{\mathbf{x}}^2(\mathbf{u};T(\mathbf{u}))\}d\mathbf{u}<+\infty, \  \forall \mathbf{x} \in G$. To fix ideas, here are two examples of orthogonal random measure which can be used to simulate the Random Function $Z$ \citep{Chi12}:
\begin{enumerate}
\item Consider a homogeneous Poisson point process on $\mathds{R}^p$ with intensity $\lambda$. Let $N(A)$ the number of points that falls in a measurable set $A \subset \mathds{R}^p$. The function $W(.)$ defined by $W(A)=N(A)-\lambda\nu(A)$ is an orthogonal random measure, where $\nu(.)$ is the $p$-dimensional Lebesgue measure.
\item Gaussian white noise on $\mathds{R}^p$ is an orthogonal random measure $W(.)$ such that: $W(A)$ is normally distributed with mean $0$ and variance $\lambda\nu(A)$; when $A$ and $B$ are disjoint $W(A)$ and $W(B)$ are independent and $W(A \cup B)=W(A) + W(B)$.
\end{enumerate}

\begin{prop}\label{Prop1}
Under the model specified in (\ref{Eq1}), we have the following moment properties:
\begin{eqnarray}{\label{Eq2}}
\mathds{E}(Z(\mathbf{x})) &=& 0, \quad \forall \mathbf{x} \in G, \label{Eq2a} \\
\mathds{C}\mbox{ov}(Z(\mathbf{x}),Z(\mathbf{y}))&=& \int_{\mathcal{T}}C(\mathbf{x},\mathbf{y};t)M(dt), \quad \forall (\mathbf{x},\mathbf{y}) \in G \times G,\label{Eq2b}\\
\mbox{where  } C(\mathbf{x},\mathbf{y};t)&=& \int_{\mathds{R}^p}f_{\mathbf{x}}(\mathbf{u};t)f_{\mathbf{y}}(\mathbf{u};t)d\mathbf{u} \quad
\mbox{and} \quad M(dt)= \lambda \mu(dt). \label{Eq2c}
\end{eqnarray}
\end{prop}

The proposition \ref{Prop1} gives a very general and flexible class of non-stationary covariance functions. This latter is defined at any two locations and is valid on $\mathds{R}^p \times \mathds{R}^p$, that is to say positive definite. Therefore, the Random Function $Z$ is non-stationary through its second order moment. The proof of the above properties is deferred to the \ref{appendix-sec1}.
 
\newpage
Note that the class of non-stationary covariance functions obtained in (\ref{Eq2b}) is a mixture of convolutions. It allows to generate different forms of regularity at the origin including low degree of regularity. Indeed, the operation of randomization by the positive measure $M(.)$ produces models that can be less regular than the base model induced by $f_{\mathbf{x}}(.;t)$, but never more regular.

\section{Construction of Closed-Form Non-stationary Covariance Functions}
\label{sec3}

Consider $\mathbf{\Sigma} : \mathds{R}^p \rightarrow PD_{p}(\mathds{R}), \mathbf{x} \mapsto \mathbf{\Sigma}_{\mathbf{x}}$ a mapping from $\mathds{R}^p$ to $PD_{p}(\mathds{R})$ the set of real-valued positive definite $p$-dimensional square matrices. Let $R^{S}(.)$ be
a continuous isotropic stationary correlation function, positive definite on $\mathds{R}^p$, for all $p \in \mathds{N}^\star$.

Let 
$\displaystyle{\phi_{\mathbf{xy}}={\left|\mathbf{\Sigma}_{\mathbf{x}}\right|}^{\frac{1}{4}}{\left|\mathbf{\Sigma}_{\mathbf{y}}\right|}^{\frac{1}{4}}{\left|{\frac{\mathbf{\Sigma}_{\mathbf{x}}+\mathbf{\Sigma}_{\mathbf{y}}}{2}}\right|}^{-\frac{1}{2}}}$ and  $\displaystyle{Q_{\mathbf{xy}}(\mathbf{h})={\mathbf{h}}^{T}{\left(\frac{\mathbf{\Sigma}_{\mathbf{x}}+\mathbf{\Sigma}_{\mathbf{y}}}{2}\right)}^{-1}\mathbf{h}}, \  \forall \mathbf{h} \in \mathds{R}^p $. 

Consider the families of functions  $\{g(.;t)\}$ and $\{k_{\mathbf{x}}(.;t), \mathbf{x} \in \mathds{R}^p\}$ for all $t \in \mathcal{T}$ such that  $\forall \mathbf{x} \in \mathds{R}^p, g(\mathbf{x};.) \in L^{2}(M)$ and  $k_{\mathbf{x}}(.;t)$ is the $p$-variate Gaussian density function centered at $\mathbf{x}$, with covariance matrix $\frac{t^2}{4}\mathbf{\Sigma}_{\mathbf{x}}$.

The following propositions and corollaries are proven in the \ref{appendix-sec1}.

\begin{prop}\label{Prop2}
If $f_{\mathbf{x}}(.;t)=g(\mathbf{x};t)k_{\mathbf{x}}(.;t)$, then the class of non-stationary covariance functions generated through \eqref{Eq2b} is:
\begin{equation}{\label{Eq3}}
C^{NS}(\mathbf{x},\mathbf{y})
=\pi^{-\frac{p}{2}}{\left| \frac{\mathbf{\Sigma}_\mathbf{x}+ \mathbf{\Sigma}_\mathbf{y}}{2} \right|}^{-\frac{1}{2}}\int_{0}^{+\infty}g(\mathbf{x};t)g(\mathbf{y};t)t^{-p}\exp\left(-\frac{Q_{\mathbf{xy}}(\mathbf{x}-\mathbf{y})}{t^2}\right)M(dt).
\end{equation}
\end{prop}

Proposition \ref{Prop2} provides a very general class of non-stationary covariance functions
similar to that proposed by \citet{Ste05}. The corollaries which will follow show that appropriate choices for the function $g(.;t)$ and the positive measure $M(.)$ give explicit classes of non-stationary covariance functions known from the literature. We thus establish a direct link between these latter and the convolution model \eqref{Eq1}.

\begin{cor}\label{Cor1}
If $g(\mathbf{x};t) \propto 1 $ and $M(dt)=t^p\xi(dt)$, for $\xi(.)$ a finite positive measure on $\mathds{R}_{+}$ such that\\
$\displaystyle{R^{S}(\tau)=\int_0^{+\infty}\exp\left(-\frac{\tau^2}{t^2}\right)\xi(dt), \tau \geq 0}$,
then the class of non-stationary correlation functions generated through \eqref{Eq3} is:
\begin{equation}{\label{Eq4}}
R^{NS}(\mathbf{x},\mathbf{y})=\phi_{\mathbf{xy}}R^{S}\left(\sqrt{ Q_{\mathbf{xy}}(\mathbf{x-y})}\right).
\end{equation}
\end{cor}

We thus find the class of closed-form non-stationary correlation functions introduced by \citet{Pac06}. The intuition behind this class is that to each input location $\mathbf{x}$ is assigned a local Gaussian kernel matrix $\mathbf{\mathbf{\Sigma}}_{\mathbf{x}}$ and the correlation between two targets $\mathbf{x}$ and $\mathbf{y}$ is calculated by averaging between the two local kernels at $\mathbf{x}$ and $\mathbf{y}$. In this way, the local characteristics at both locations influence the correlation of the corresponding target values. Thus, it is possible to account for non-stationarity. It is done by specifying the mapping $\mathbf{\Sigma}(.)$ which models the anisotropy of the correlation function. The resulting kernel matrix $\mathbf{\mathbf{\Sigma}}_{\mathbf{x}}$ at each point $\mathbf{x}$ is interpreted as a locally varying geometric anisotropy matrix. It controls the anisotropic behavior of the Random Function in a small neighborhood around $\mathbf{x}$.

The positive measure  $\xi(.)$ defined in the corollary \ref{Cor1}  is the measure associated with the Schoenberg representation of the correlation function $R^{S}(.)$ which is a continuous, positive definite and radial function on $\mathds{R}^p$, for all $p \in \mathds{N}^\star$ \citep{Sch88}. An approach based on a class of non-stationary correlation models defined in \eqref{Eq4} does not allow the use of isotropic stationary correlation models that are not valid in all dimensions. Thus, we lost  the isotropic stationary correlation functions with compact support such that the spherical model (valid in $\mathds{R}^p, \ p\leq 3$).

\newpage
\begin{example}\label{Ex1}
By a specific choice of the positive measure $M(.)$ in corollary \ref{Cor1}, we obtain some subclasses of closed-form non-stationary correlation functions:

\begin{enumerate}
\item $M(dt)=t^p\delta_a(t)\mathds{1}_{[0,+\infty)}(t)dt, a>0, \ R^{NS}(\mathbf{x},\mathbf{y})=\phi_{\mathbf{xy}}\exp\left(-\frac{Q_{\mathbf{xy}}(\mathbf{x-y})}{a^2}\right)$;
\item $M(dt)=\frac{t^p}{a\sqrt{\pi}}\exp\left(-\frac{t^2}{4a^2}\right)\mathds{1}_{[0,+\infty)}(t)dt, a>0, \ R^{NS}(\mathbf{x},\mathbf{y})=\phi_{\mathbf{xy}}\exp\left(-\frac{\sqrt{ Q_{\mathbf{xy}}(\mathbf{x-y})}}{a}\right)$;
\item $M(dt)=2t^{p+1}h(t^2)\mathds{1}_{[0,+\infty)}(t)dt$, with $h(.)$ the density of the Gamma distribution $\mathcal{G}a (\nu,1/4a^2), \nu, a>0,
R^{NS}(\mathbf{x},\mathbf{y})=\phi_{\mathbf{xy}}\frac{2^{1-\nu}}{\Gamma(\nu)}\left({\frac{\sqrt{Q_{\mathbf{xy}}(\mathbf{x-y})}}{a}}\right)^{\nu}K_{\nu}\left({\frac{\sqrt{Q_{\mathbf{xy}}(\mathbf{x-y})}}{a}}\right)$, where $K_{\nu}(.)$ is the modified Bessel function of second kind with order $\nu$ \citep{Gra07};
\item $M(dt)=2t^{p+1}h(t^2)\mathds{1}_{[0,+\infty)}(t)dt$, with $h(.)$ the density of the inverse Gamma distribution $\mathcal{IG} (\alpha,a^2), \alpha, a>0,\  R^{NS}(\mathbf{x},\mathbf{y})=\phi_{\mathbf{xy}}{\left(1+\frac{Q_{\mathbf{xy}}(\mathbf{x-y})}{a^2}\right)}^{-\alpha}.$
\end{enumerate}
\end{example}

The example \ref{Ex1} shows that a suitable choice of the positive measure $M(.)$ produces non-stationary versions of some well known stationary correlation functions: gaussian, exponential, Matérn and Cauchy. These examples can also be deduced from the following corollaries \ref{Cor2} and \ref{Cor3}.

\begin{cor}\label{Cor2}
If $g(\mathbf{x};t) \propto t^{\nu(\mathbf{x})}, \nu(\mathbf{x})>0, \forall \mathbf{x} \in \mathds{R}^p$ and $ M(dt)=2t^{p-1}h(t^2)\mathds{1}_{[0,+\infty)}(t)dt$, with $h(.)$ the density of  Gamma distribution $\mathcal{G}a(1,1/4a^2), a>0$, then the class of non-stationary correlation functions generated through \eqref{Eq3} is:
\begin{equation}{\label{Eq5}}
R^{NS}(\mathbf{x},\mathbf{y})=\phi_{\mathbf{xy}}\frac{2^{1-\nu(\mathbf{x},\mathbf{y})}}{\sqrt{\Gamma(\nu(\mathbf{x}))\Gamma(\nu(\mathbf{y}))}}M_{\nu(\mathbf{x},\mathbf{y})}\left(\sqrt{ Q_{\mathbf{xy}}(\mathbf{x-y})}\right),
\end{equation}
where $M_{\nu(\mathbf{x},\mathbf{y})}(\tau)={\left(\tau/a\right)}^{\nu(\mathbf{x},\mathbf{y})} K_{\nu(\mathbf{x},\mathbf{y})}\left(\tau/a\right),\  \tau \geq 0$; $K_{\nu(\mathbf{x},\mathbf{y})}(.)$  is the modified Bessel function of second kind with order  $\nu(\mathbf{x},\mathbf{y})=(\nu(\mathbf{x})+\nu(\mathbf{y}))/2$.
\end{cor}

From corollary \ref{Cor2}, we retrieve the non-stationary version of the stationary Matérn correlation function introduced by \citet{Ste05}. This one  allows both local geometric anisotropy and degree of differentiability to vary spatially. When the function $\nu(.,.)$ is constant, we obtain the non-stationary Matérn correlation function with constant regularity parameter shown in example \ref{Ex1}. However, the associated function  $g(;t)$ and positive measure $M(.)$ are different.

\begin{cor}\label{Cor3}
If $g(\mathbf{x};t) \propto t^{-\alpha(\mathbf{x})}, \alpha(\mathbf{x})>0, \forall \mathbf{x} \in \mathds{R}^p$ et $ M(dt)=2t^{p+3}h(t^2)\mathds{1}_{[0,+\infty)}(t)dt$, with $h(.)$ the density of  inverse Gamma distribution $\mathcal{IG} (1,a^2), a>0$, then the class of non-stationary correlation functions generated through \eqref{Eq3} is:
\begin{equation}{\label{Eq6}}
\displaystyle{
R^{NS}(\mathbf{x},\mathbf{y})=\phi_{\mathbf{xy}}\frac{\Gamma(\alpha(\mathbf{x},\mathbf{y}))}{\sqrt{\Gamma(\alpha(\mathbf{x}))\Gamma(\alpha(\mathbf{y}))}}{\left(1+\frac{Q_{\mathbf{xy}}(\mathbf{x-y})}{a^2}\right)}^{-\alpha(\mathbf{x},\mathbf{y})}},
\end{equation}
where $\displaystyle{\alpha(\mathbf{x},\mathbf{y})=(\alpha(\mathbf{x})+\alpha(\mathbf{y}))/2}$.
\end{cor}

The corollary \ref{Cor3} gives  a non-stationary version of the Cauchy correlation function introduced by \citet{Ste05}. Both the local geometric anisotropy and long-range dependence are allowed to vary spatially. For the constant long-range dependence parameter $\alpha(.,.)$, we obtain the Cauchy non-stationary correlation function shown in example \ref{Ex1}.

From above formulas, it is straightforward to construct a closed-form non-stationary covariance function including a standard deviation function. Doing so, the non-stationary covariance is defined as follows:
\begin{equation}\label{Eq7}
C^{NS}(\mathbf{x},\mathbf{y})
=\sigma(\mathbf{x})\sigma(\mathbf{y})R^{NS}(\mathbf{x},\mathbf{y}),
\end{equation}
where $\sigma(.)$ is a standard deviation function  and $R^{NS}(.,.)$ a closed-form non-stationary correlation function.

Note that, the closed-form non-stationary covariance functions defined in \eqref{Eq7} have the desirable property of including stationary covariance functions as a special case. 

\section{Inference}
\label{sec4}

In this section, we propose a procedure for estimation of spatially variable parameters that govern the classes of closed-form non-stationary second order structure model presented in Section  \ref{sec3}. Although it could be used for all of these classes, we focus on the one defined in \eqref{Eq4} which is flexible enough for many applications. The proposed estimation methodology allows us to deal simultaneously with other types of non-stationarity: mean and variance. Thus, we consider an extended model built from the model specified in \eqref{Eq1} which includes a non-stationarity both in mean and variance. The estimation procedure is based on the local stationarity assumption and is achieved using a three-step estimation scheme. First, a non-parametric kernel estimator  of the local variogram is built. Then, it is used in a weighted local least squares procedure to estimate parameters at a representative set of points referred to as anchor points. Next, a kernel smoothing approach is used to interpolate the parameters at any location of interest. 

\subsection{Modelling}
\label{ssec1}

To include a non-stationarity both in mean and variance, we consider a Random Function $Y=\{Y(\mathbf{x}): \mathbf{x} \in G \subseteq \mathds{R}^p , p \geq 1\}$ described as follows:

\begin{equation}\label{Eq9}
Y(\mathbf{x})=m(\mathbf{x})+ \sigma(\mathbf{x})Z(\mathbf{x}), \ \forall  \mathbf{x} \in G,
\end{equation}
where $m: \mathds{R}^p \rightarrow \mathds{R}$ is an unknown fixed function. $\sigma: \mathds{R}^p \rightarrow \mathds{R}^{+}$ is an unknown positive fixed function. $Z$ is a zero-expectation, unit variance Random Function with correlation function defined in \eqref{Eq4}:
\begin{equation*}
R^{NS}(\mathbf{x},\mathbf{y})=\phi_{\mathbf{xy}}R^{S}\left(\sqrt{ Q_{\mathbf{xy}}(\mathbf{x-y})}\right).
\end{equation*}

Thus, the Random Function $Z$ carries the spatial dependence structure of the Random Function $Y$. The model formulation defined in \eqref{Eq9} leads to a first and second order moments written as follows:
\begin{eqnarray}\label{Eq10}
\mathds{E}(Y(\mathbf{x})) 
&=& m(\mathbf{x}),\label{Eq10a} \\
\mathds{C}\mbox{ov}(Y(\mathbf{x}),Y(\mathbf{y}))
&=&\sigma(\mathbf{x})\sigma(\mathbf{y})R^{NS}(\mathbf{x},\mathbf{y})\equiv C^{NS}(\mathbf{x},\mathbf{y}) \label{Eq10b}.
\end{eqnarray}

From the expressions (\ref{Eq10a}) and (\ref{Eq10b}), we see that the non-stationarity of the Random Function $Y$  is  characterized by  the parameters $\sigma(.)$, $\mathbf{\Sigma}(.)$ and $m(.)$ defined at any location of the domain of interest.

Let $\mathbf{Y}={(Y(\mathbf{s}_1),\ldots,Y(\mathbf{s}_n))}^T$ be a $(n \times 1)$ vector of observations from a unique realization of the Random Function $Y$, associated to known locations $\{\mathbf{s}_1,\ldots,\mathbf{s}_n\}\subset G\subseteq \mathds{R}^p$. The goal is to use the data $\mathbf{Y}$ to infer the standard deviation function $\sigma(.)$, the correlation function determined by $\mathbf{\Sigma}(.)$ and the mean function $m(.)$. Next, we use them to predict the value of the Random Function $Y$ at unsampled locations. From now on, without loss of generality, we assume that $p=2$.

\subsection{Parameter Estimation}
\label{ssec2}
The estimation of the parameters $\sigma(.)$, $\mathbf{\Sigma}(.)$ and $m(.)$ relies on the quasi-stationarity or local stationarity assumption which  allows certain simplifications. Specifically, we work with the following definition of local stationarity introduced by \citet{Mat71}.
\newpage
\subsubsection{Local Stationarity}
\label{sssec1}
 A Random Function $Y=\{Y(\mathbf{x}): \mathbf{x} \in G \subseteq \mathds{R}^p , p \geq 1\}$ will be locally stationary if it has an expectation $m(\mathbf{x})$ and covariance $C^{NS}(\mathbf{x},\mathbf{y})$ such that:
\begin{enumerate}
\item $m(\mathbf{x})$ is a very regular function varying slowly  in space at the scale of the available information; more precisely, $m(\mathbf{x})$ can be considered as constant in a neighborhood of $\mathbf{x}$;
\item There is a function of three arguments $C^S(\mathbf{h};\mathbf{x},\mathbf{y})$ such that $C^{NS}(\mathbf{x},\mathbf{y})=C^S(\mathbf{x-y};\mathbf{x},\mathbf{y})$ and such that, for a given $\mathbf{h}$, $C^S(\mathbf{h};\mathbf{x},\mathbf{y})$ is a very regular and slowly varying (in the same sense as in 1.) function of the two arguments $\mathbf{x}$ and $\mathbf{y}$. In other words, for locations $\mathbf{x}$ and $\mathbf{y}$ not too far from each other, $C^S(\mathbf{h};\mathbf{x},\mathbf{y})$ only depends on $\mathbf{h}$, as is if the covariance $C^{NS}(.,.)$ was  stationary.
\end{enumerate}

The intuitive idea behind this definition is that if a Random Function is locally stationary, then at any location $\mathbf{x}_0 \in G$ there exists a neighborhood $\mathcal{V}_{\mathbf{x}_0}$ where the Random Function can be approximated by a stationary Random Function. Thus,
$\forall (\mathbf{x}, \mathbf{y}) \in  \mathcal{V}_{\mathbf{x}_0} \times \mathcal{V}_{\mathbf{x}_0},\  m(\mathbf{x}) \approx m(\mathbf{y}) \approx m(\mathbf{x}_0)  \mbox{  and   } C(\mathbf{x},\mathbf{y}) \approx C^S(\mathbf{x}-\mathbf{y};\mathbf{x}_0)=C^S(\mathbf{h};\mathbf{x}_0), \ \|\mathbf{h}\| \leq b$; where $C^S(.)$ is a stationary covariance function and the limit $b$ represents the radius of the quasi-stationarity neighborhood $\mathcal{V}_{\mathbf{x}_0}$. In this way, the parameters  are assumed  to be very smooth functions which vary slowly over the domain. The expectation of the Random Function being approximately equal to a constant inside the quasi-stationarity neighborhood, the resulting local covariance structure at any location $\mathbf{x}_0$ is written as follows:
\begin{equation}{\label{Eq11}}
C^S(\mathbf{h};\mathbf{x}_0)= {\sigma}^2({\mathbf{x}_0})R^S\left(\sqrt{{\mathbf{h}}^{T}{\mathbf{\Sigma}}_{\mathbf{x}_0}^{-1}\mathbf{h}}\right), \ \|\mathbf{h}\| \leq b.
\end{equation}

In more practical terms, we can define moving neighborhoods  $\mathcal{V}_{\mathbf{x}_0}=\{\mathbf{x} \in G, \ \|\mathbf{x}-\mathbf{x}_0\| \leq b\}$ within which the expectation and the covariance can be considered stationary, and where the available information is sufficient to make the inference. 

The quasi-stationarity assumption is a compromise between the distances of homogeneity of the studied phenomenon and the density of the available information. Indeed, it is always possible to reach the stationarity by reducin the size $b$ of neighborhoods. But most of these neighborhoods will contain almost no data; therefore inference of parameters will not be possible in these neighborhoods.

\subsubsection{Anisotropy Parameterization}
\label{sssec2}

Locally, the non-stationary spatial dependence structure defined in \eqref{Eq10b} is thus reduced to an anisotropic stationary one \eqref{Eq11}. The anisotropy function ${\mathbf{\Sigma}}(.)$ is parameterized using the spectral decomposition. By this way, the positive definiteness is guaranteed and the locally varying  geometric anisotropy can be captured. Precisely, at any location $\mathbf{x}_0 \in G$,  ${\mathbf{\Sigma}}_{\mathbf{x}_0}={\mathbf{\Psi}}_{\mathbf{x}_0}{\mathbf{\Lambda}}_{\mathbf{x}_0}{\mathbf{\Psi}}_{\mathbf{x}_0}^T$, where ${\mathbf{\Lambda}}_{\mathbf{x}_0}$ is the diagonal matrix of eigenvalues and ${\mathbf{\Psi}}_{\mathbf{x}_0}$ is the eigenvector matrix. Then, we have

$
{\mathbf{\Lambda}}_{\mathbf{x}_0}=
\begin{pmatrix}
\lambda_1^2({\mathbf{x}_0})&0\\
0&\lambda_2^2({\mathbf{x}_0})
\end{pmatrix}
,
\quad
{\mathbf{\Psi}}_{\mathbf{x}_0}=
\begin{pmatrix}
\cos\psi({\mathbf{x}_0})&\sin\psi({\mathbf{x}_0})\\
-\sin\psi({\mathbf{x}_0})&\cos\psi({\mathbf{x}_0})
\end{pmatrix}
,
\quad
\lambda_1({\mathbf{x}_0}),\lambda_2({\mathbf{x}_0})>0$ and $\psi({\mathbf{x}_0}) \in [0,\pi).
$ 

Doing so, the scale parameter of the isotropic stationary correlation function $R^S(.)$ in the expression \eqref{Eq11} is set to one to avoid any overparameterization. At each point, the square roots of the eigenvalues control the local ranges and the eigenvector matrix specify the local orientations. Thus, the anisotropy function ${\mathbf{\Sigma}}(.)$ is characterized by the functions  $\lambda_1(.)$, $\lambda_2(.)$ and $\psi(.)$.

\newpage
\subsubsection{Local Variogram Kernel Estimator}
\label{sssec3}

It is convenient to estimate the second order spatial structure through the variogram. Under the local stationarity framework, we define a non-parametric kernel moment estimator of the stationary local  variogram at a fixed location $\mathbf{x}_0 \in G$  and lag $\mathbf{h} \in \mathds{R}^p$, $\gamma(\mathbf{h};\mathbf{x}_0)={\sigma}^2({\mathbf{x}_0})-C^S(\mathbf{h};\mathbf{x}_0), \ \|\mathbf{h}\| \leq b$  as follows:
\begin{equation}\label{Eq12}
{\widehat{\gamma}}_{\epsilon}(\mathbf{h};\mathbf{x}_0)=\frac{\sum_{V(\mathbf{h})}K^{\star}_{\epsilon}(\mathbf{x}_0,\mathbf{s}_i)K^{\star}_{\epsilon}(\mathbf{x}_0,\mathbf{s}_j){[Y(\mathbf{s}_i)-Y(\mathbf{s}_j)]}^2}{2\sum_{V(\mathbf{h})}K^{\star}_{\epsilon}(\mathbf{x}_0,\mathbf{s}_i)K^{\star}_{\epsilon}(\mathbf{x}_0,\mathbf{s}_j)}, \ \|\mathbf{h}\| \leq b,
\end{equation}
where the average \eqref{Eq12} is taken over $V(\mathbf{h})=\{(\mathbf{s}_i,\mathbf{s}_j): \mathbf{s}_i-\mathbf{s}_j = \mathbf{h} \}$, the set of all pairs of locations separated by vector $\mathbf{h}$; $K^{\star}_\epsilon(\mathbf{x}_0,\mathbf{s}_i)=K_\epsilon(\mathbf{x}_0,\mathbf{s}_i)/\sum_{l=1}^nK_\epsilon(\mathbf{x}_0,\mathbf{s}_l)$ are standardized weights; $K_\epsilon(.,.)$ is a non-negative, symmetric  kernel on $\mathds{R}^p \times \mathds{R}^p$ with bandwidth parameter $\epsilon >0$. For irregularly spaced data there, are generally not enough observations separated by exactly $\mathbf{h}$. Then, $V(\mathbf{h})$ is usually modified to $\{(\mathbf{s}_i,\mathbf{s}_j): \mathbf{s}_i-\mathbf{s}_j \in T(\mathbf{h}) \}$, where $T(\mathbf{h})$ is a tolerance region of $\mathds{R}^p$ surrounding $\mathbf{h}$.

This moment estimator of the local variogram at any location $\mathbf{x}_0 \in G$ is a kernel weighted local average of squared differences of the regionalized variable.  The kernel function is used to smoothly downweight the squared differences (for each lag interval) according to the distance of these paired values from a target location. We assign to each data pair a weight proportional to the product of the individual weights. Observation pairs near to the target location $\mathbf{x}_0$ have more influence on the local variogram estimator than those which are distant.

Note that when $K_{\epsilon}(\mathbf{x},\mathbf{y})\propto 1, \forall (\mathbf{x},\mathbf{y}) \in G \times G $ and $b=D/2$; where $D$ is the diameter of the domain of interest $G$, we obtain the classical Matheron moment estimator for a global stationary structure   \citep{Mat71}. For $K_{\epsilon}(\mathbf{x},\mathbf{y}) \propto \mathds{1}_{\|\mathbf{x}-\mathbf{y}\|<\epsilon}, \forall (\mathbf{x},\mathbf{y}) \in G \times G$ and $b=\epsilon$, we get the moving window estimator  \citep{Haa90b}. 

To calculate the non-parametric kernel estimator \eqref{Eq12}, we opt for an isotropic stationary Gaussian kernel: $K_{\epsilon}(\mathbf{x},\mathbf{y})\propto \exp(-\frac{1}{2\epsilon^2}{\| \mathbf{x}-\mathbf{y}\|}^2), \ \forall (\mathbf{x},\mathbf{y}) \in G \times G$. The latter has a non-compact support and therefore considers all observations. Thus, the local variogram estimator is not limited only to the local information, distant points are also considered. This avoids artifacts  caused by the only use of observations close to the target location. It also reduces instability of the obtained local variogram at regions with low sampling density. Furthermore, it  provides a smooth parameter estimate and then is compatible with the quasi-stationarity assumption. Concerning the size of the quasi-stationarity neighborhood $b$, it is set with respect to the bandwidth $\epsilon$. We take $b=\sqrt{3}\epsilon$ such that the standard deviation of the isotropic stationary Gaussian kernel  match the isotropic stationary uniform kernel (with compact support). Another possible choice for $b$ is to take a quantile of the isotropic stationary Gaussian kernel (e.g. $b\approx 2\epsilon$) or full width at half maximum ($b=\sqrt{2\log(2)}\epsilon$).

\subsubsection{Raw Estimates of Parameters}
\label{sssec4}

We want to estimate the vector of structural parameters $\boldsymbol{\theta}(.)=\left({\sigma}(.),\lambda_1(.),\lambda_2(.),\psi(.)\right)$ and the mean parameter $m(.)$ at any location of the domain of interest.

The estimation of the parameters vector $\boldsymbol{\theta}(\mathbf{x}_0)$ which characterizes the stationary local  variogram $\gamma(.;\mathbf{x}_0)\equiv \gamma(.;\boldsymbol{\theta}(\mathbf{x}_0)) $ at a fixed location $\mathbf{x}_0$ are found via the following  minimization problem:
\begin{eqnarray}{\label{Eq13}}
\widehat{\boldsymbol{\theta}}(\mathbf{x}_0)=\argmin_{\theta(\mathbf{x}_0) \in \Theta} \| \mathbf{w}_{\epsilon}(\mathbf{x}_0)\odot(\boldsymbol{\gamma}(\boldsymbol{\theta}(\mathbf{x}_0))-\widehat{\boldsymbol{\gamma}}_{\epsilon}(\mathbf{x}_0))\|,
\end{eqnarray}
where $\odot$ is the product term by term ; $\boldsymbol{\gamma}(\boldsymbol{\theta}(\mathbf{x}_0))={[\gamma(\mathbf{h}_1;\boldsymbol{\theta}(\mathbf{x}_0)),\ldots,\gamma(\mathbf{h}_{J};\boldsymbol{\theta}(\mathbf{x}_0))]}^T$;  $\widehat{\boldsymbol{\gamma}}_{\epsilon}(\mathbf{x}_0)={[\widehat{\gamma}_{\epsilon}(\mathbf{h}_1;\mathbf{x}_0),\ldots,\widehat{\gamma}_{\epsilon}(\mathbf{h}_{J};\mathbf{x}_0)]}^T$; $\mathbf{w}_{\epsilon}(\mathbf{x}_0)={[w_{\epsilon}(\mathbf{h}_1;\mathbf{x}_0),\ldots,w_{\epsilon}(\mathbf{h}_{J};\mathbf{x}_0)]}^T$, $w_{\epsilon}(\mathbf{h};\mathbf{x}_0)={\left[(\sum_{V(\mathbf{h})}K^{\star}_{\epsilon}(\mathbf{x}_0,\mathbf{s}_i)K^{\star}_{\epsilon}(\mathbf{x}_0,\mathbf{s}_j))/\|\mathbf{h}\|\right]}^{1/2}$; $\{\mathbf{h}_{j} \in \mathds{R}^p, j=1,\ldots,J\}$ are given lag vectors; $\boldsymbol{\theta}(\mathbf{x}_0) \in \Theta $ is the vector of unknown parameters and $\Theta$ is an open parameter space. 

\newpage
The estimation of the structural parameters ${\sigma}(.),\lambda_1(.),\lambda_2(.)$ and $\psi(.)$ depends on the bandwidth parameter $\epsilon$ through the kernel function $K_{\epsilon}(.)$. The bandwidth parameter controls the range of validity of the stationary approximation and its selection is adresssed in Section \ref{ssec3}. Note that the estimation of the structural parameters does not require the prior estimation of the mean function $m(.)$. Moreover, no model is specified for this latter.

Concerning the estimation of the parameter $m(.)$, the mean $m(\mathbf{x}_0)$ at a fixed location $\mathbf{x}_0$ is approximatively equal to a constant inside the quasi-stationarity neighborhood $\mathcal{V}_{\mathbf{x}_0}=\{\mathbf{x} \in G, \ \|\mathbf{x}-\mathbf{x}_0\| \leq b\}$. Thus, using the estimate of the vector of structural parameters $\widehat{\boldsymbol{\theta}}(\mathbf{x}_0)$ obtained in \eqref{Eq13}, $m(\mathbf{x}_0)$ is estimated explicitly by a local stationary kriging of the mean \citep{Mat71}. Specifically we have:
\begin{equation}{\label{Eq14}}
\widehat{m}(\mathbf{x}_0)= \sum_{\mathbf{s}_i \in \mathcal{V}_{\mathbf{x}_0} }\alpha_i({\mathbf{x}_0})Y(\mathbf{s}_i),
\end{equation}
where  $\boldsymbol{\alpha}_{\mathbf{x}_0}=[\alpha_i({\mathbf{x}_0})]$ are the kriging weights given by:
\begin{equation}{\label{Eq15}}
\boldsymbol{\alpha}_{\mathbf{x}_0}=\frac{\mathbf{\widehat{\Gamma}}_{\mathbf{x}_0}^{-1}\mathbf{1}}{\mathbf{1}^T \mathbf{\widehat{\Gamma}}_{\mathbf{x}_0}^{-1}\mathbf{1}},
\end{equation}
with $\mathbf{\widehat{\Gamma}}_{\mathbf{x}_0}=[\gamma(\mathbf{s}_i - \mathbf{s}_j;\widehat{\boldsymbol{\theta}}(\mathbf{x}_0))], (\mathbf{s}_i, \mathbf{s}_j) \in \mathcal{V}_{\mathbf{x}_0} \times \mathcal{V}_{\mathbf{x}_0}$.

\subsubsection{Smoothing Parameters}
\label{sssec5}

For the spatial prediction purpose, one needs to compute the parameters  $\sigma(.)$, $\mathbf{\Sigma}(.)$ and $m(.)$ at prediction and observation locations. In practice, it is unnecessary to solve the minization problem \eqref{Eq13} at each target location. Indeed, doing so is computationally intensive and redundant for close locations, since these estimates are highly correlated. To reduce the computational burden, the proposed idea consists in obtainning the parameter estimates only at some reduced set of $m \ll n$ representative points  referred to as anchor points defined over the domain. Then, using the estimates obtained at anchor points, a kernel smoothing method is used to make available estimates at any location of interest. Since the parameters ${\sigma}(.),\lambda_1(.),\lambda_2(.),\psi(.)$ and $m(.)$ are supposed to vary slowly in space (quasi-stationarity), the Nadaraya-Watson kernel estimator seems appropriate, in addition to being relatively simple. However, other smoothers can be used as well (local polynomials, splines, \ldots).

The Nadaraya-Watson kernel estimator \citep{Wan95} of ${\sigma}(.)$ at any location ${\mathbf{x}_0} \in G$ is given by:
\begin{equation}{\label{Eq16}}
\widetilde{{\sigma}}({\mathbf{x}_0})=\sum_{k=1}^m W_{k}(\mathbf{x}_0)\widehat{\sigma}(\mathbf{x}_k), \quad W_{k}(\mathbf{x}_0)=\frac{K(\frac{\mathbf{x}_0-\mathbf{x}_k}{\delta})}{{\sum_{k=1}^m K(\frac{\mathbf{x}_0-\mathbf{x}_k}{\delta})}}, 
\end{equation}
where  $K(.)$ is a kernel on $\mathds{R}^p$; $\delta>0$ is a smoothing parameter; $\{\widehat{\sigma}(\mathbf{x}_k),k=1,\ldots,m\}$ are the raw estimates of parameter ${\sigma}(.)$ at anchor points $\{\mathbf{x}_k, k=1,\ldots,m\}$.

We similarly define kernel smoothing estimator $\widetilde{\lambda}_1(.)$, $\widetilde{\lambda}_2(.)$ and $\widetilde{m}(.)$ at any location ${\mathbf{x}_0} \in G$. For the specific case of the orientation parameter $\psi(.)$, the kernel estimator at any location ${\mathbf{x}_0} \in G$ is given through the minimization of the following criteria:
\begin{equation}{\label{Eq17}}
\widetilde{\psi}(\mathbf{x}_0)=\argmin_{\psi_0\in S^p}\sum_{k=1}^m W_{k}(\mathbf{x}_0){d^2(\psi_0,\widehat{\psi}(\mathbf{x}_k))},
\end{equation}
where  $\{\widehat{\psi}(\mathbf{x}_k), k=1,\ldots,m\}$ are the raw estimates of parameter  $\psi(.)$ at anchor points $\{\mathbf{x}_k, k=1,\ldots,m\}$; $S^p$ is the $p$-dimensional sphere of unit radius its centre at the origin; $d(\psi_0,\widehat{\psi}(\mathbf{x}_k))$ is a distance between two orientations. For $p=2$, we can take the $d(\psi_0,\widehat{\psi}(\mathbf{x}_k))=\min(|\psi_0-\widehat{\psi}(\mathbf{x}_k)|,|\psi_0-\widehat{\psi}(\mathbf{x}_k) - \pi|, |\psi_0-\widehat{\psi}(\mathbf{x}_k)+ \pi| )$.

Since, the parameters $\sigma(.)$, $\mathbf{\Sigma}(.)$ and $m(.)$ are assumed to vary slowly and regularly across the domain of interest (local stationarity), we choose $K(.)$ as an isotropic stationary Gaussian kernel. The selection of its smoothing parameter $\delta$ is discussed in Section \ref{ssec3}.

\subsection{Choice of Hyper-Parameters}
\label{ssec3}

In the proposed method, a crucial point is the determination of the bandwidth parameter $\epsilon$ used in the computation of the local variogram non-parametric kernel estimator  defined in \eqref{Eq12}. Indeed, the size of the local  stationarity neighborhood is expressed in terms of the bandwidth parameter. We are also interest in the choice of the bandwidth used to smooth the parameter estimates.

Concerning the bandwidth parameter $\epsilon$, in some applications it may be appropriate to choose it subjectively, but in general, it is desirable to have available methods for choosing it automatically from the data. The data-driven method used to select the bandwidth $\epsilon$ consists of leaving out one data location and using a form of cross-validation. Because the estimation of the spatial dependence structure is rarely a goal per se but an intermediate step before kriging, we want to choose the bandwidth that gives the best cross-validation mean square error (MSE). More explicitly, we minimize with respect to $\epsilon$ the leave-one-out cross-validation score \citep{Zha10}:
\begin{equation}{\label{Eq18}}
MSE(\epsilon)  =  \frac{1}{n}\sum_{i=1}^{n}{\left(Y(\mathbf{s}_{i}) - \widehat{Y}_{-i}(\mathbf{s}_{i};\epsilon)\right)}^2,
\end{equation}
where $\widehat{Y}_{-i}(\mathbf{s}_{i};\epsilon)$ denotes the spatial predictor computed at location $\mathbf{s}_{i}$ using all observations except $\{Y(\mathbf{s}_{i})\}$. The prediction method is described in Section \ref{ssec4}.

The choice of the smoothing bandwidth $\delta$ associated to the parameter  $\sigma(.)$ is done using the following cross-validation criteria \citep{Wan95}:
\begin{equation}{\label{Eq19}}
CV(\delta)=\frac{1}{m}\sum_{k=1}^m {\left(\frac{\widehat{\sigma}(\mathbf{x}_k)-\widetilde{\sigma}(\mathbf{x}_k)}{1-W_{k}(\mathbf{x}_k)}\right)}^2, 
\end{equation}
where $\{\widehat{\sigma}(\mathbf{x}_k), \widetilde{\sigma}(\mathbf{x}_k), k=1,\ldots,m\}$ are respectively the raw and smoothed estimates of the parameter $\sigma(.)$ at anchor points $\{\mathbf{x}_i, k=1,\ldots,m\}$. The cross-validation score function is minimized over a grid of smoothing bandwidths to choose the optimal value.

Theoretically, smoothing bandwidths associated to each local parameter can be different. Our numerical examples indicate that choosing the same smoothing bandwidth for all local parameters in order to reduce the computational burden, makes little difference in terms of prediction performance.  This remark was already highlighted by \citet{Zhu10}.

\section{Prediction}
\label{sec5}

The main purposes of modelling and estimating the spatial dependence structure is to spatially interpolate data  and perform conditional simulations. The expected benefit using the closed-form non-stationary covariance model \eqref{Eq10b}  is to obtain spatial predictions and variance estimation errors more realistic than those based on a inadequate stationary covariance. In this section, a description of kriging and conditional simulations based on the non-stationary model \eqref{Eq9} is presented.

\subsection{Kriging}
\label{ssec4}

Let $C^{NS}(. , .)$ the non-stationary covariance function of the Random Function $Y$ and  $m(.)$ its mean. Given the vector of observtaions $\mathbf{Y}={\left(Y(\mathbf{s}_1),\ldots,Y(\mathbf{s}_n)\right)}^T$ at $n$ fixed locations $\mathbf{s}_1,\ldots,\mathbf{s}_n \in G$, the point predictor for the unknown value of $Y$ at unsampled location $\mathbf{s}_0 \in G$ is given by the optimal linear predictor:
\begin{equation}{\label{Eq20}}
\widehat{Y}(\mathbf{s}_{0})= m(\mathbf{s}_{0}) + \sum_{i=1}^{n}\eta_{i}(\mathbf{s}_{0})(Y(\mathbf{s}_{i})-m(\mathbf{s}_{i})). 
\end{equation}

The kriging weight vector $\boldsymbol{\eta}=[\eta_i(\mathbf{s}_{0})]$ and the corresponding kriging variance $Q(\mathbf{s}_{0})$ are given by:
\begin{equation}{\label{Eq21}}
\boldsymbol{\eta}={\mathbf{C}}^{-1}\mathbf{C}_0  \quad \mbox{et } \quad Q(\mathbf{s}_{0})=\sigma
^2(\mathbf{s}_{0})- {\mathbf{C}}_0^T\mathbf{C}^{-1}\mathbf{C}_0.
\end{equation}
where $\mathbf{C}_0=[C^{NS}(\mathbf{s}_{i},\mathbf{s}_{0})]$; $\mathbf{C}={[C^{NS}(\mathbf{s}_{i},\mathbf{s}_{j})]}$.

\subsection{Conditional Simulations}
\label{ssec5}

Despite its optimality, the estimator obtained by kriging smoothes reality. Thus, the spatial texture of the reality (captured by the covariance or variogram) can not be reproduced from data points only. However, in some situations, it is more important to reproduce the spatial variability than obtain the best accuracy. We want to have a realization which has the same degree of variability than the actual phenomenon and coincides with observations at data points. To do this, we use conditional simulations. 

Here we assume that $Y$ is a Gaussian Random Function, with mean $m(.)$ and non-stationary covariance structure $C^{NS}(.,.)$. We want to simulate at a large number of locations a Gaussian Random Function with same mean and covariance, and ensure that the realization honors the observed values $Y(\mathbf{s}_1),\ldots,Y(\mathbf{s}_n)$. This can be accomplished from an unconditional simulation of the Random Function $Y$ as follows \citep{Lan02}:
\begin{enumerate}
\item realize a unconditional simulation $\{X(\mathbf{s}), \mathbf{s} \in G\}$ of the Random Function $Y$;
\item carried out a simple kriging of $\{X(\mathbf{s})-Y(\mathbf{s}), \mathbf{s} \in G\}$ from its values taken at the data points $\{\mathbf{s}_i, i=1,\ldots,n\}$, using $m(.)$ and $C^{NS}(.,.)$ ;
\item add the unconditional simulation and the result of kriging.
\end{enumerate}

We have $Y(\mathbf{x})=m(\mathbf{x})+ \sigma(\mathbf{x})Z(\mathbf{x}), \forall \mathbf{x} \in G$, where $Z$ is a Gaussian Random Function with zero expectation, unit variance and non-stationary correlation function $R^{NS}(.,.)$. Thus, to simulate the Gaussian Random Function $Y$ (step 1 of the previous algorithm), we need to known how we can simulate $Z$. Simulation of the Gaussian Random Function $Z$ can be carried out using a propagative version of the Gibbs sampler proposed by \citet{Lan12}. This algorithm allows to simulate a Gaussian vector at a large number of locations (comparatively to the existing classical algorithms such as Cholesky method or Gibbs sampler) without relying on a Markov assumption (it does not need to have a sparse precision matrix).  The algorithm proposed in \citep{Lan12} requires  neither the inversion nor the factorization of a covariance matrix.  Note that simulation methods such as spectral method or turning bands method are not adapted to the non-stationary case \citep{Lan02}. The representation  that underlies these methods relies on the stationarity assumption. Formally, the algorithm proposed by \citet{Lan12} is outlined below:

Let $\mathbf{Z}=(Z_a,a \in A \subset G)$ be a standardized Gaussian vector, with covariance matrix $\mathbf{C}=(C_{\alpha\beta},\alpha,\beta \in A)$. The underlying idea is that instead of simulating the vector $\mathbf{Z}$, we can simulate the vector $\mathbf{X}=\mathbf{C}^{-1}\mathbf{Z}$. This latter is a Gaussian vector with covariance matrix $\mathbf{C}^{-1}$. Noting that the inverse of $\mathbf{C}^{-1}$ is precisely $\mathbf{C}$, the Gibbs sampler is applied to $\mathbf{X}$ and $\mathbf{Z}$ is updated accordingly. This gives the following algorithm:
\begin{enumerate}
\item set $z^c_G=0$;
\item generate $a \sim {\cal U}(G)$;
\item generate $z^n_a \sim \mathcal{N}(0,1)$ and take $z^n_b=z^c_b+C_{ab}(z^n_a-z^c_a)$ for each $b\neq a$;
\item take $z^c_G=z^n_G$ and go to 2.
\end{enumerate}

Thus, at each step of the algorithm, a Gaussian value is assigned randomly to a pivot, and then spread to the other components.

\newpage
\section{Applications}
\label{sec6}

In this section, we illustrate the advantages of our proposed approach and its potential on two datasets: a soil data and a rainfall data. A comparison scheme of kriging under stationary and non-stationary models is carried out through a validation sample. Next, we compare prediction and  prediction standard deviation maps on a fine grid. The aim is to compare the behaviour of the models in terms of spatial interpolation and associated uncertainty.

\subsection{Soil Data Example}
\label{ssec6}
We now study a dataset coming from a soil gamma radiometric potassium concentration (in shots per second, cps) survey in the region of the Hunter Valley, NSW, Australia. This data set was used by \citet{McB13} to illustrate their approach. We have a training data (537 observations) which serves to calibrate the model and a validation data (1000 observations) which serves only to assess the prediction performances. 

\begin{figure}[h!]
        \centering
        \begin{subfigure}[H]{0.35\textwidth}
                \centering
                \includegraphics[width=1\textwidth]{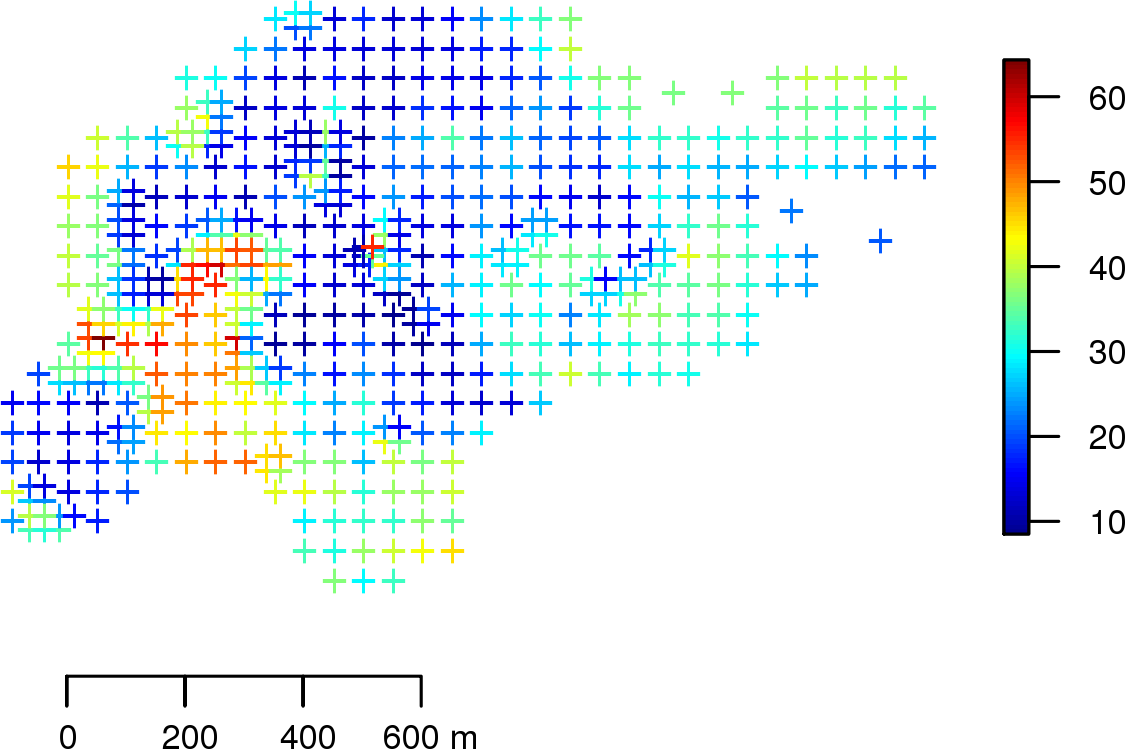}
                \caption{}\label{Fig1a}
        \end{subfigure}
         \qquad     
        \begin{subfigure}[H]{0.35\textwidth}
                \centering
                \includegraphics[width=1\textwidth]{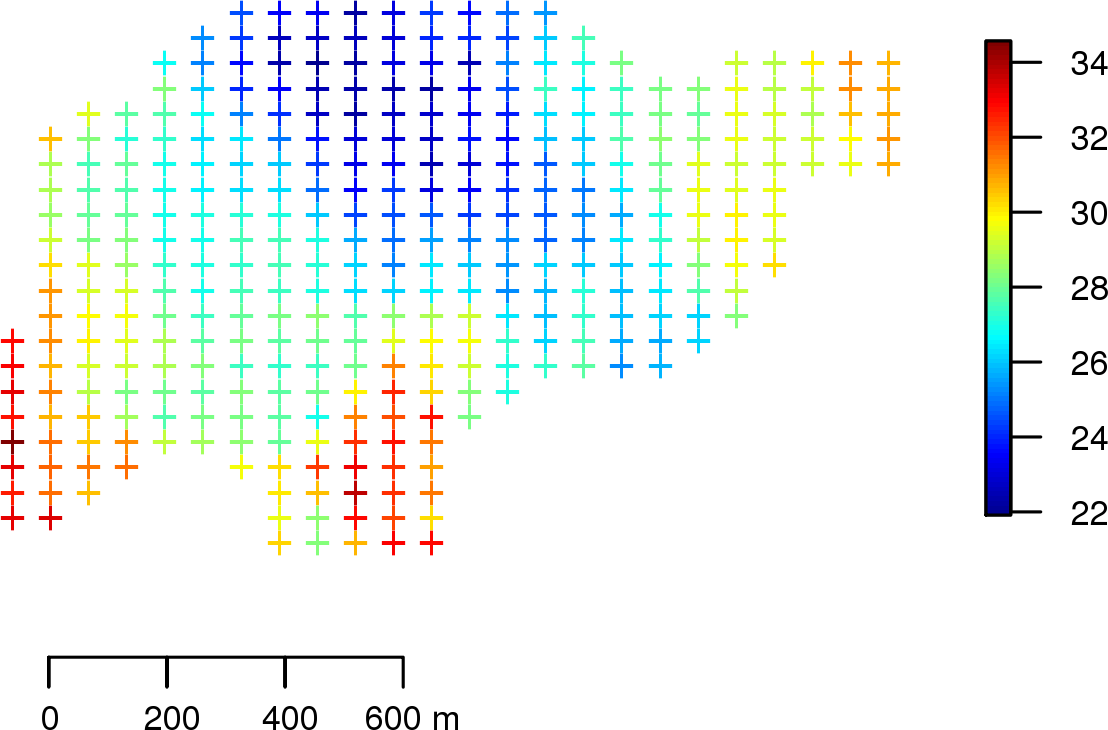}
                \caption{}\label{Fig1b}
        \end{subfigure}
        
        \medskip 
        \begin{subfigure}[H]{0.35\textwidth}
                \centering
                \includegraphics[width=1\textwidth]{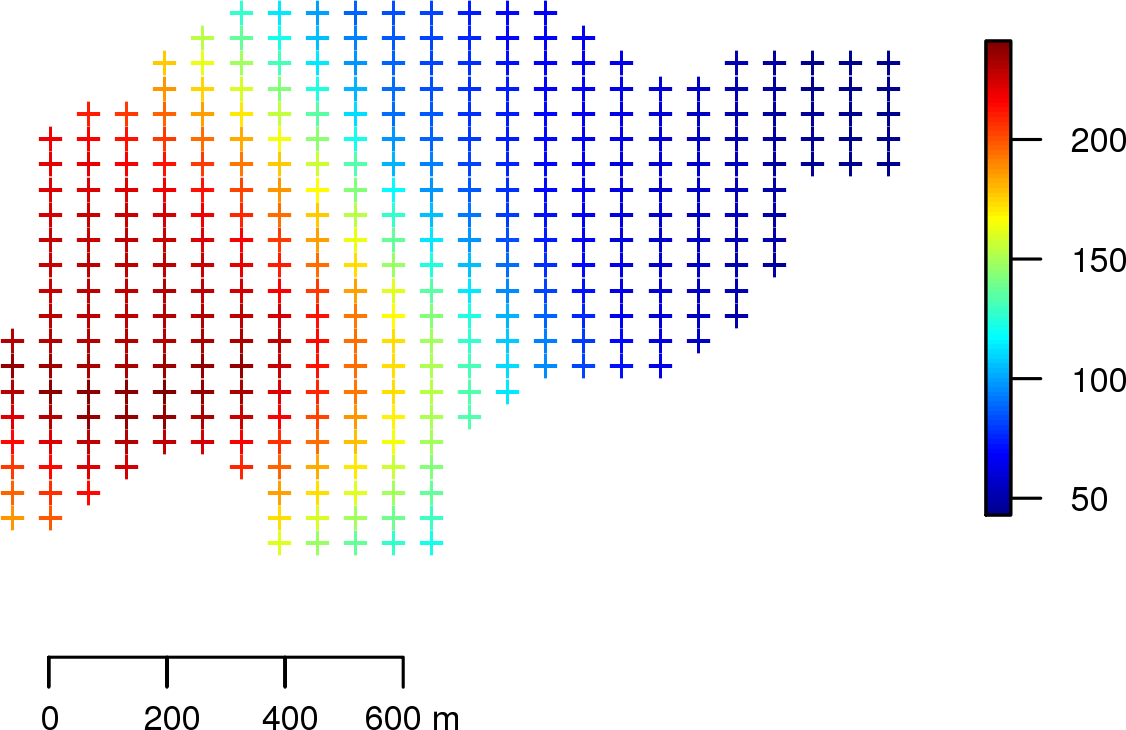}
                \caption{}\label{Fig1c}%
        \end{subfigure}
        \qquad 
        \begin{subfigure}[H]{0.35\textwidth}
                \centering
                \includegraphics[width=1\textwidth]{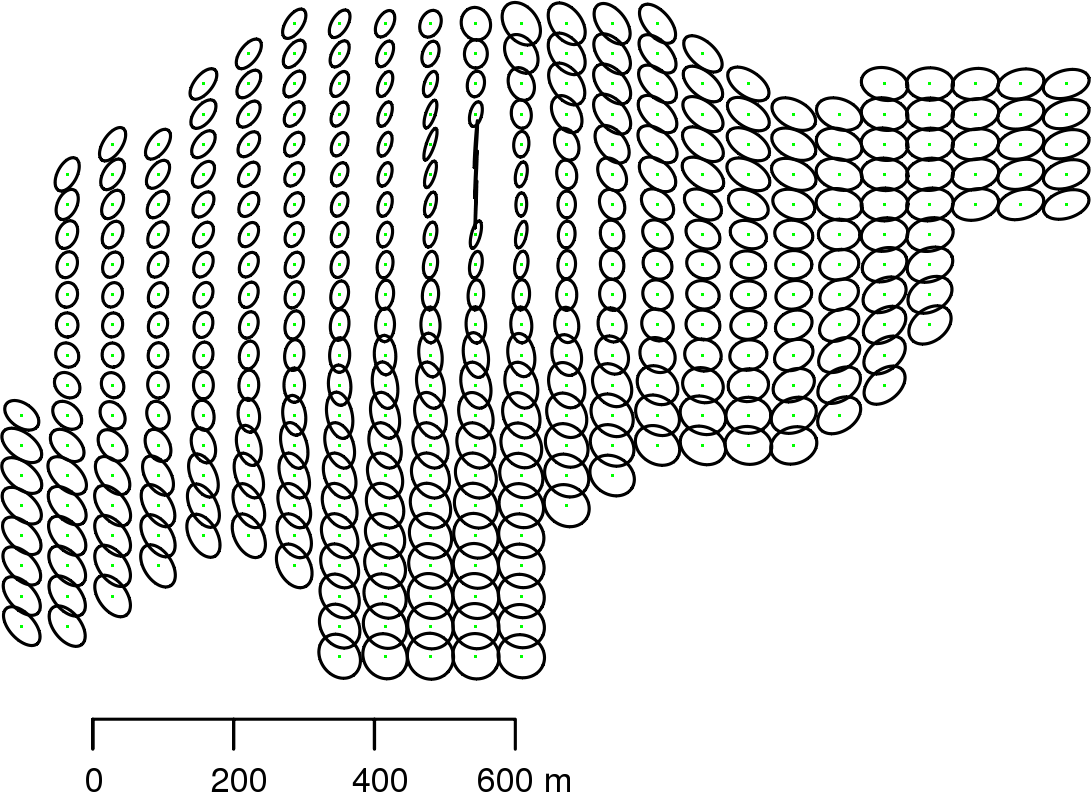}
                \caption{}\label{Fig1d}
        \end{subfigure}
        \caption{(a) Training data; (b) Estimated mean function $\widehat{\mbox{m}}(.)$ at anchor points; (c) Estimated variance function $\widehat{\sigma}^2(.)$ at anchor points;  (d) Estimated anisotropy function $\widehat{\mathbf{\Sigma}}(.)$ at anchor points where the ellipses were scaled to ease vizualisation. (Potassium concentration data)\label{Fig1}}
\end{figure}

Raw estimates of parameters $m(.)$, $\sigma^2(.)$ and $\mathbf{\Sigma}(.)$ at anchor points are shown respectively on Figures \ref{Fig1b}, \ref{Fig1c} and \ref{Fig1d}. They are based on the non-stationary exponential covariance function (Example \ref{Ex1}). Concerning the estimated anisotropy function $\widehat{\mathbf{\Sigma}}(.)$ at anchor points, it is represented by ellipses as shown in Figure \ref{Fig1d}. Indeed, covariance (positive definite) matrices have an appealing geometrical interpretation: they can be uniquely identified with an ellipsoid. Based on these estimates, non-stationarity in the data is quite visible. Especially, from Figure \ref{Fig1d} where we can clearly see the spatially varying azimuth. Such directional effects are also quite apparent on data (Figure \ref{Fig1a}). Note that the stationary approach has not detected a global geometric anisotropy.

\newpage

\begin{figure}[h!]
        \centering
        \begin{subfigure}[h!]{0.35\textwidth}
                \centering
                \includegraphics[width=1\textwidth]{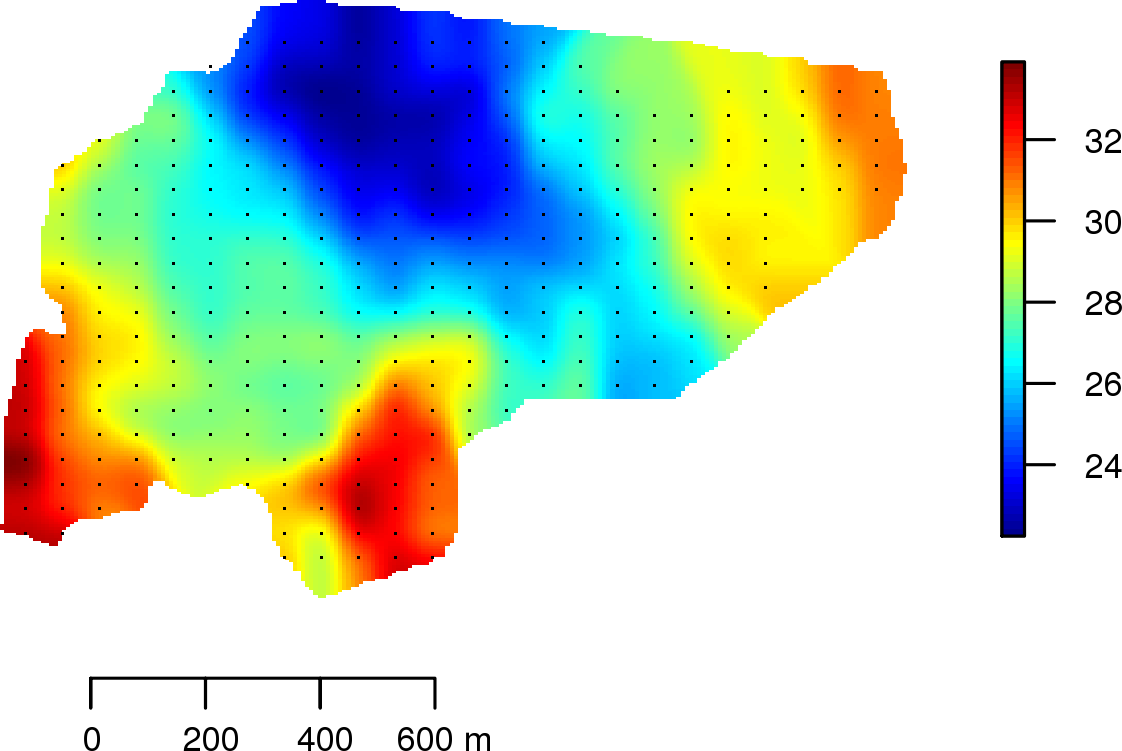}
                \caption{}\label{Fig2a}
        \end{subfigure}
        \qquad 
        \begin{subfigure}[h!]{0.35\textwidth}
                \centering
                \includegraphics[width=1\textwidth]{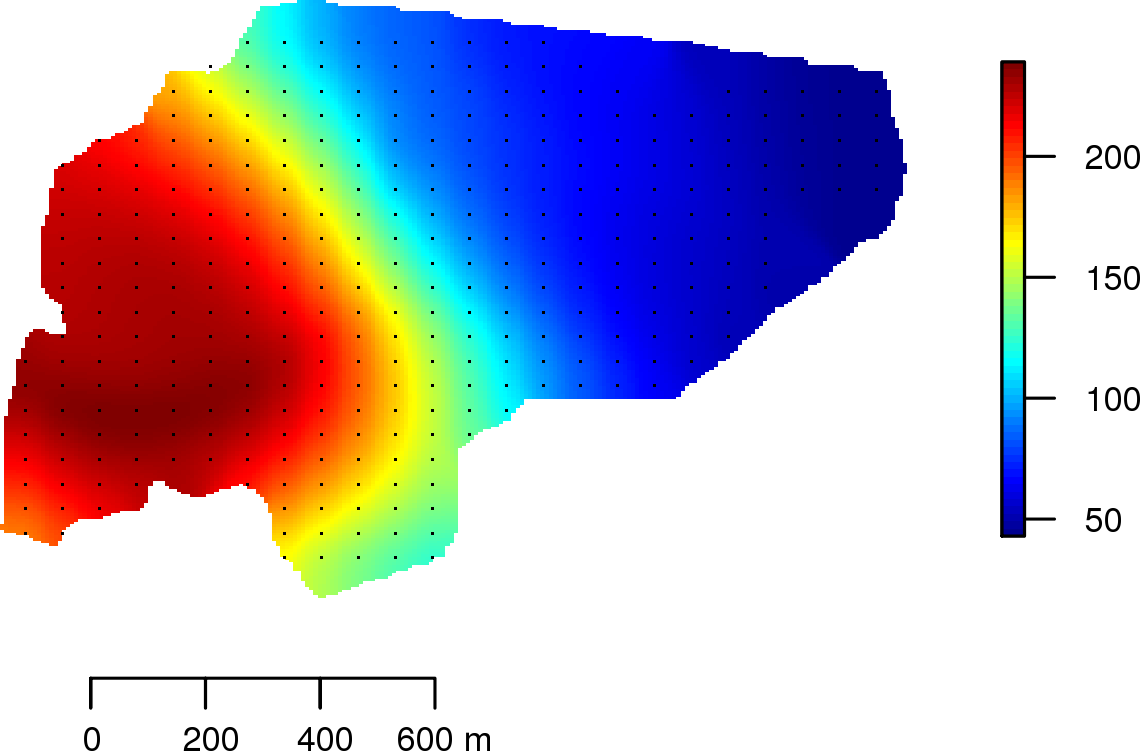}
                \caption{}\label{Fig2b}
        \end{subfigure}
        
        \medskip
        
        \begin{subfigure}[h!]{0.35\textwidth}
                \centering
                \includegraphics[width=1\textwidth]{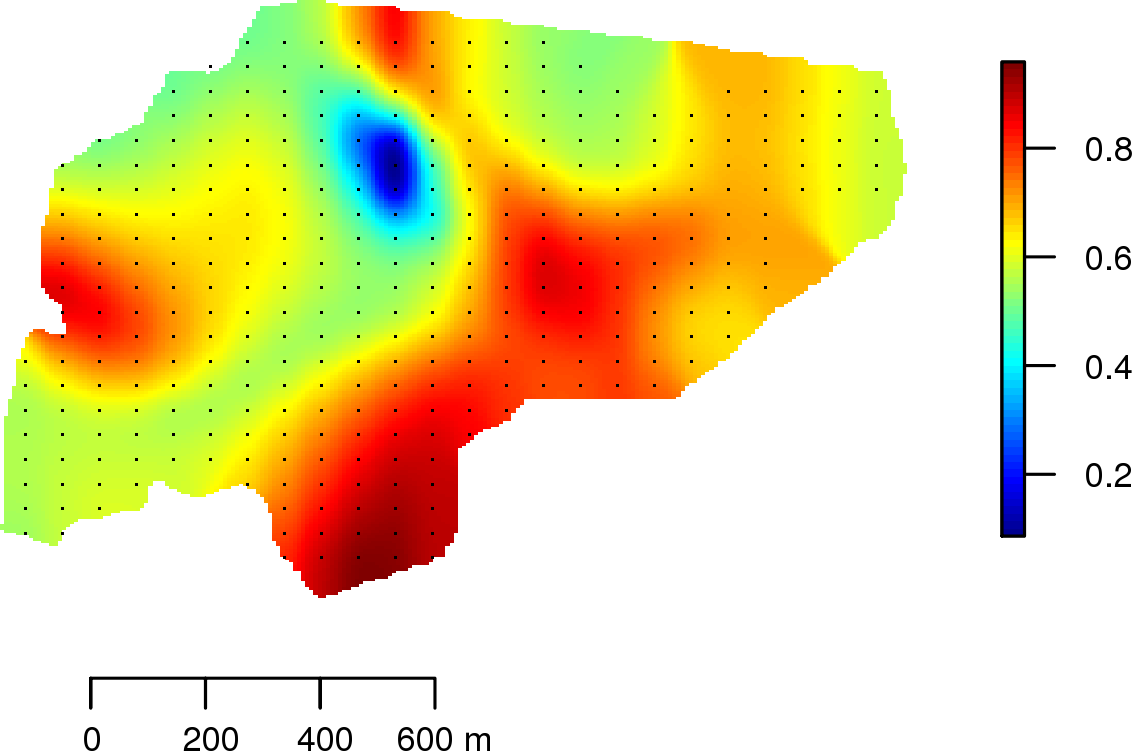}
                \caption{}\label{Fig2c}
        \end{subfigure}
        \qquad 
        \begin{subfigure}[h!]{0.35\textwidth}
                \centering
                \includegraphics[width=1\textwidth]{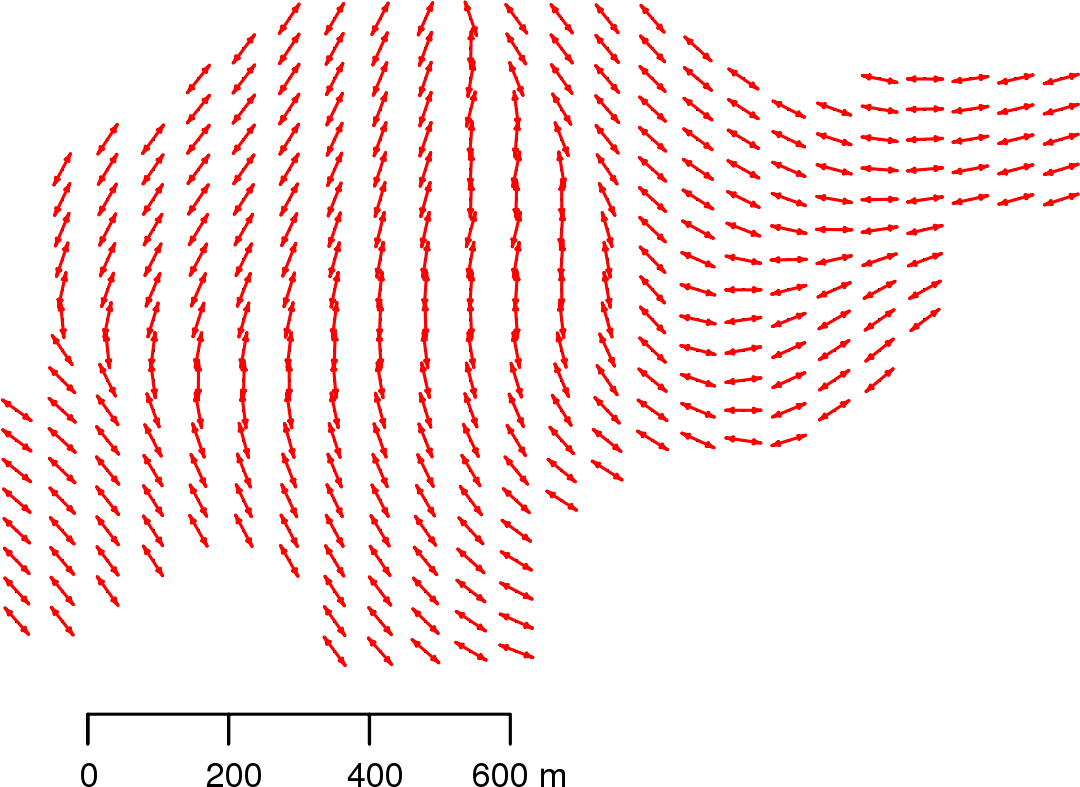}
                \caption{}\label{Fig2d}
        \end{subfigure}
        \caption{Smoothed parameters over the domain of observations: (a) mean, (b) variance, (c) anisotropy ratio, (d) azimuth. (Potassium concentration data)}\label{Fig2}
\end{figure}

Following the hyper-parameters selection procedure presented in Section \ref{ssec3}, the bandwidth associated with non-parametric Gaussian kernel estimator of the local variogram is $\epsilon=224$ m. Figure \ref{Fig2} shows the maps of smoothed parameters over the whole domain of observations: mean, variance, anisotropy ratio and azimuth. The optimal smoothing bandwidth associated to the Gaussian kernel smoothing corresponds to $\delta=30$ m, following the selection procedure described in Section \ref{ssec3}.

\begin{figure}[h!]
        \centering
        \begin{subfigure}[h]{0.35\textwidth}
                \centering
                \includegraphics[width=1\textwidth]{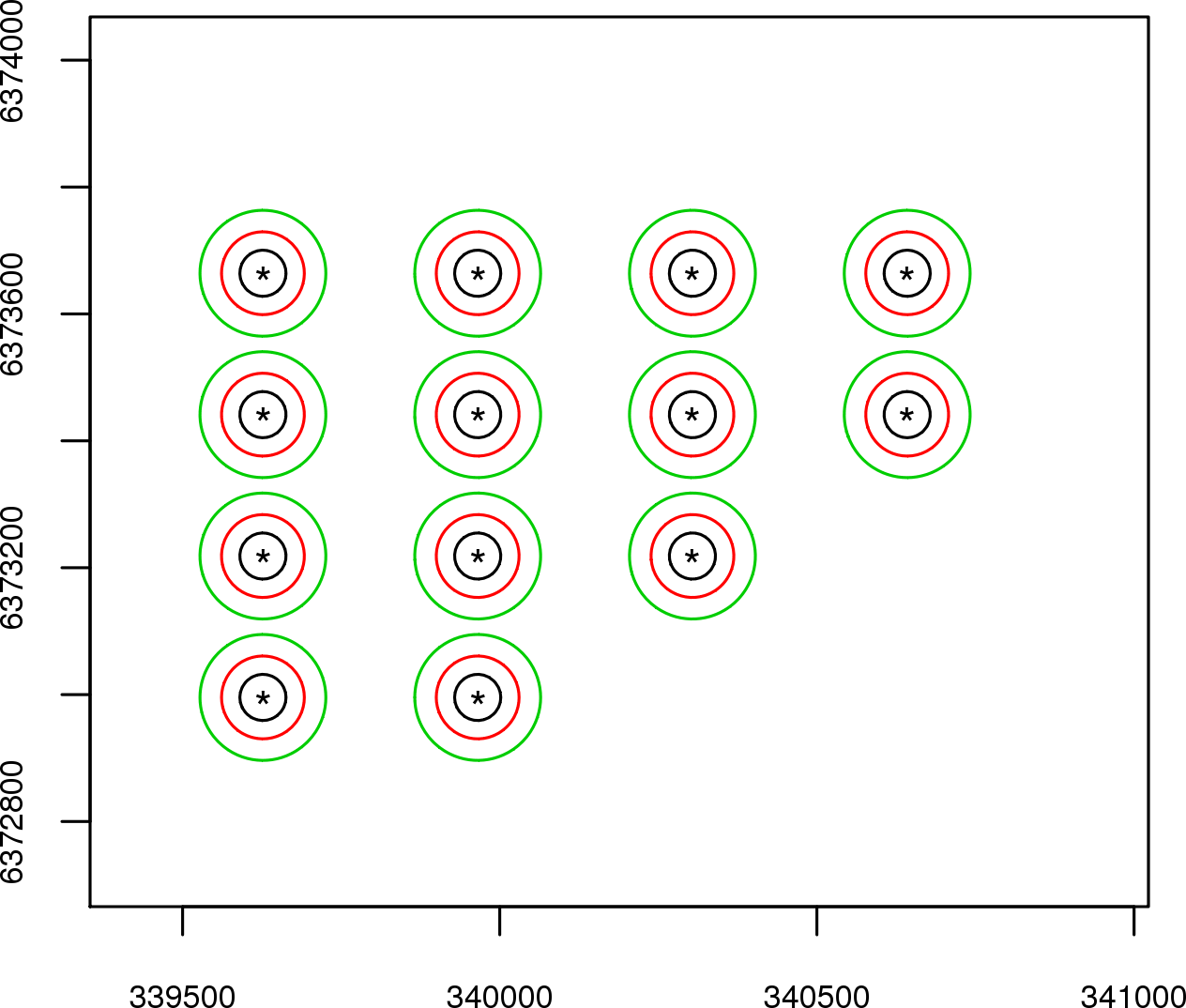}
                \caption{}\label{Fig3a}
        \end{subfigure}
        \qquad
        \begin{subfigure}[h]{0.35\textwidth}
                \centering
                {\includegraphics[width=1\textwidth]{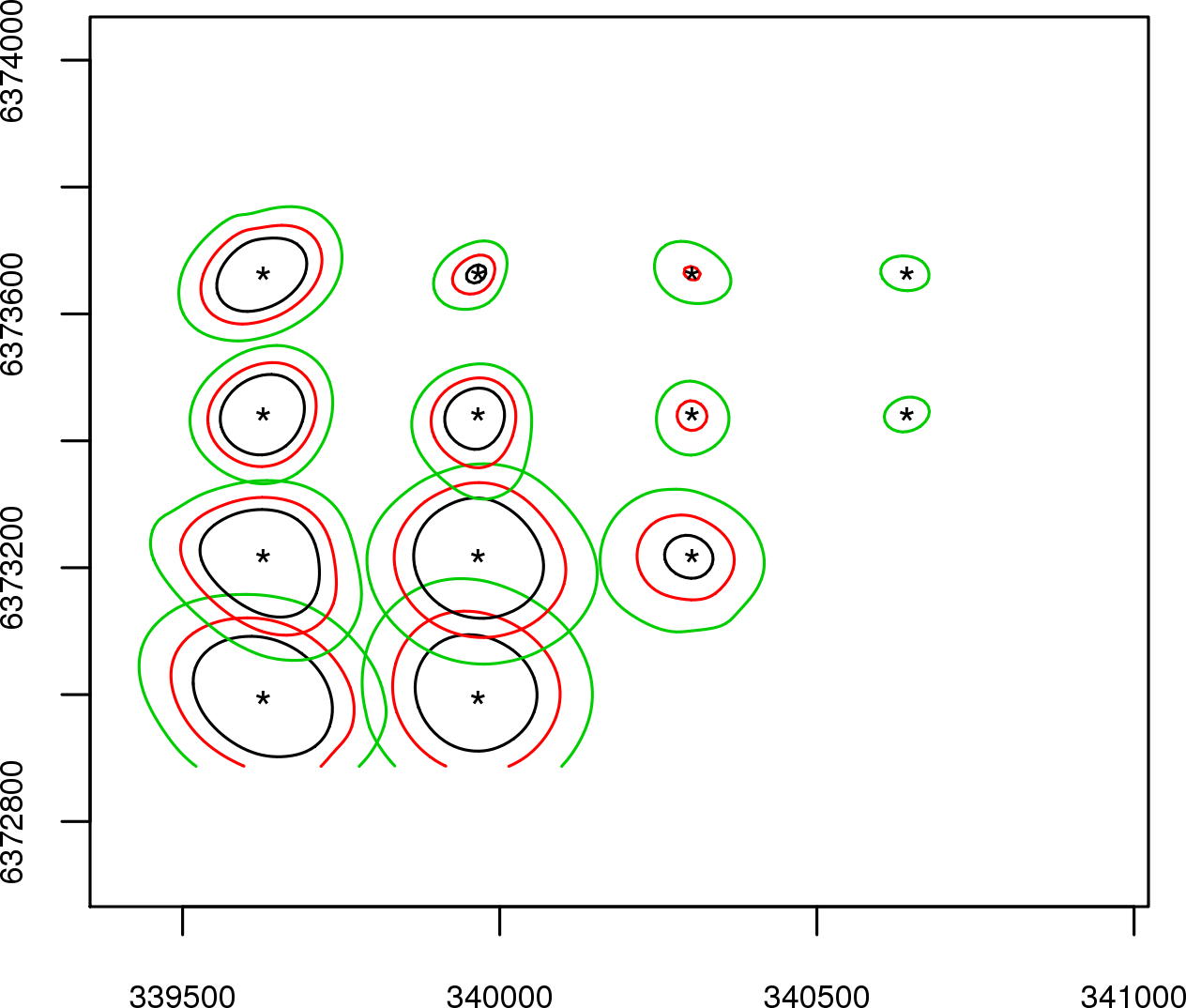}}
                \caption{}\label{Fig3b}
        \end{subfigure}
        \caption{Covariance  level contours at few points for the estimated stationary and non-stationary models (a, b). Level contours correspond to the values: $80$ (black), $60$ (red) and $40$ (green). (Potassium concentration data)}\label{Fig3}
\end{figure}

A visualization of the covariance at certain points for estimated stationary and non-stationary models is presented in Figure \ref{Fig3}. We can see how the non-stationary spatial dependence structure changes the shape from one place to another as compared to the stationary one. The stationary model is a nested isotropic model (nugget effect, exponential and spherical) while the non-stationary model corresponds to the non-stationary exponential covariance function (Example \ref{Ex1}). 

\begin{figure}[h!]
        \centering
        \begin{subfigure}[h!]{0.35\textwidth}
                \centering
                \includegraphics[width=1\textwidth]{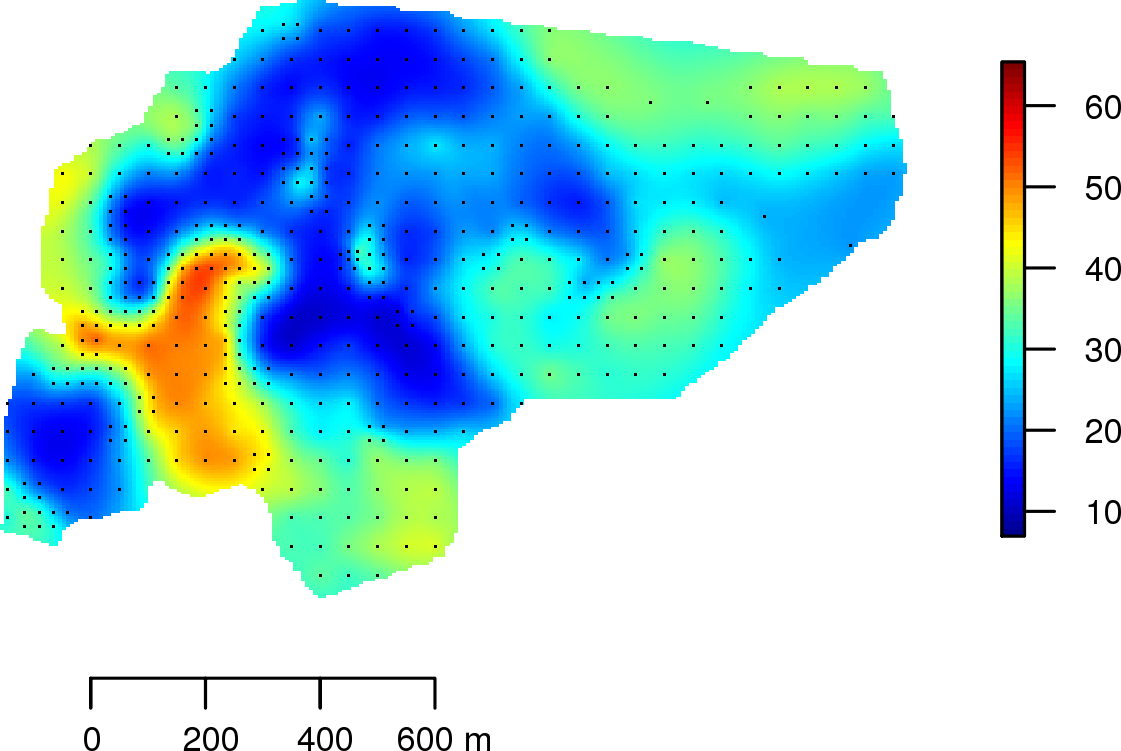}
                \caption{}\label{Fig4a}
        \end{subfigure}
        \qquad 
        \begin{subfigure}[h!]{0.35\textwidth}
                \centering
                \includegraphics[width=1\textwidth]{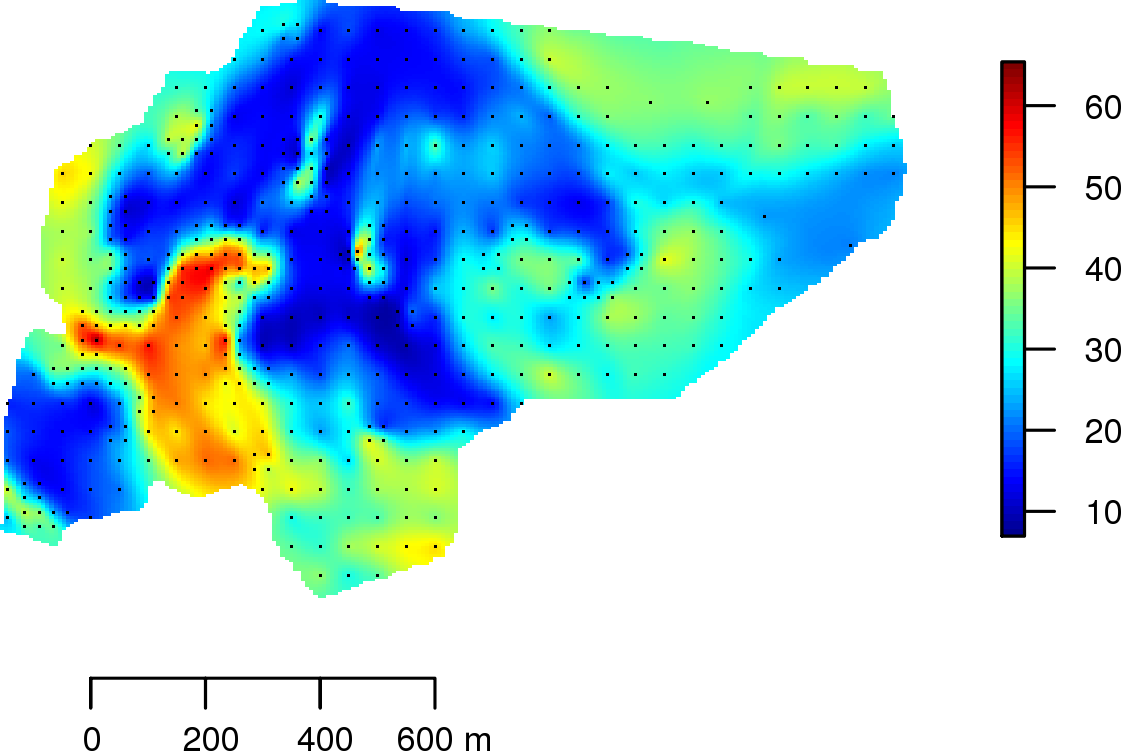}
                \caption{}\label{Fig4b}
        \end{subfigure}
        
        \medskip
        
        \begin{subfigure}[h!]{0.35\textwidth}
                \centering
                \includegraphics[width=1\textwidth]{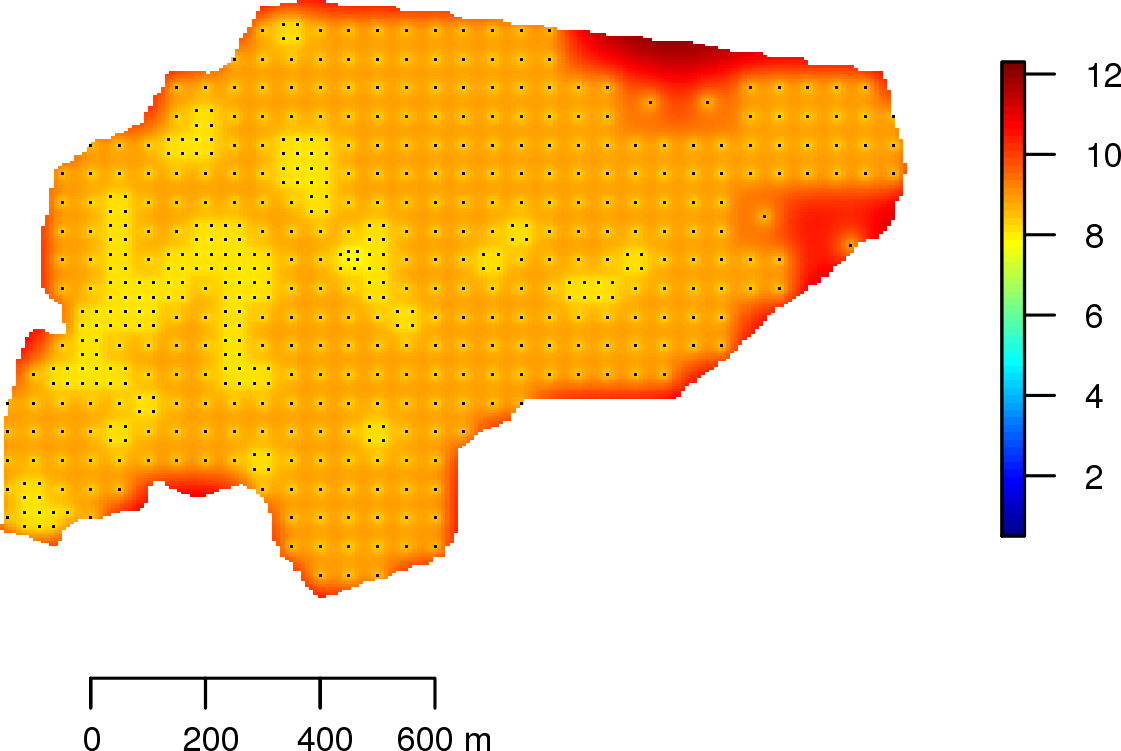}
                \caption{}\label{Fig4c}
        \end{subfigure}
        \qquad 
        \begin{subfigure}[h!]{0.35\textwidth}
                \centering
                \includegraphics[width=1\textwidth]{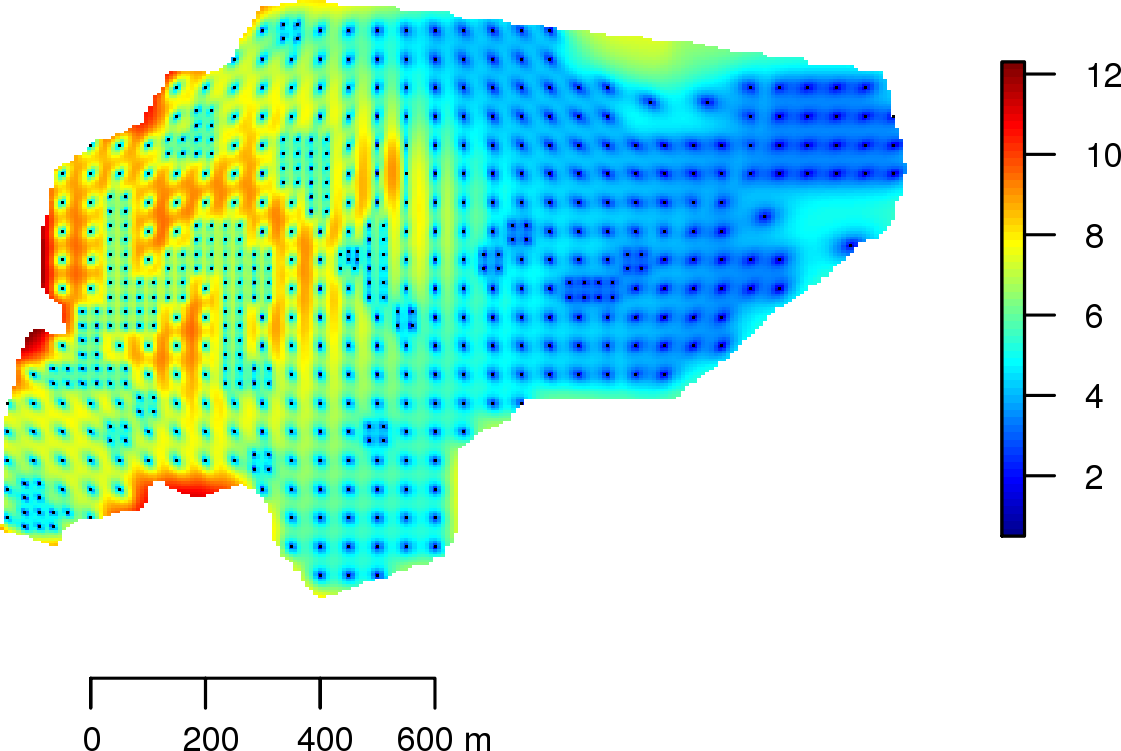}
                \caption{}\label{Fig4d}
        \end{subfigure}
        \caption{(a,b) Predictions and prediction standard deviations for the estimated stationary model. (c,d) Predictions and prediction standard deviations for the estimated non-stationary model. (Potassium concentration data)}\label{Fig4}
\end{figure}

\begin{figure}[h!]
        \centering      
        \begin{subfigure}[h]{0.35\textwidth}
                \centering
                \includegraphics[width=1\textwidth]{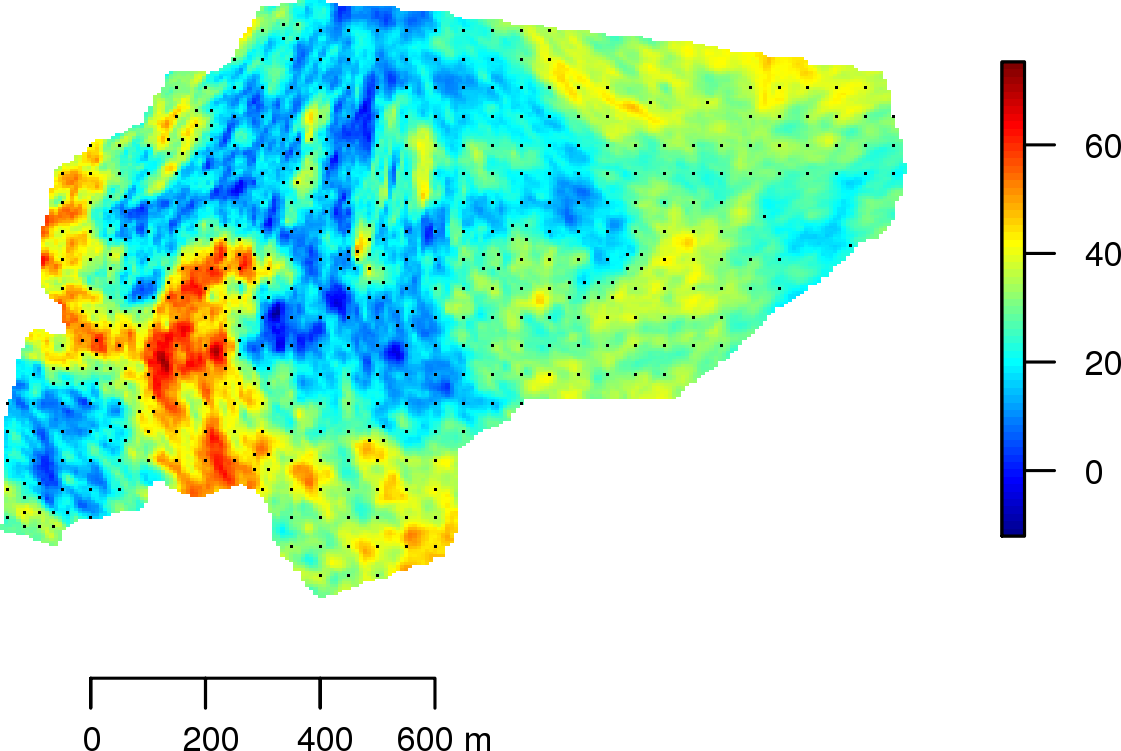}
                \caption{Simulation \#1}\label{Fig5a}
        \end{subfigure}
         \qquad    
        \begin{subfigure}[h]{0.35\textwidth}
                \centering
                \includegraphics[width=1\textwidth]{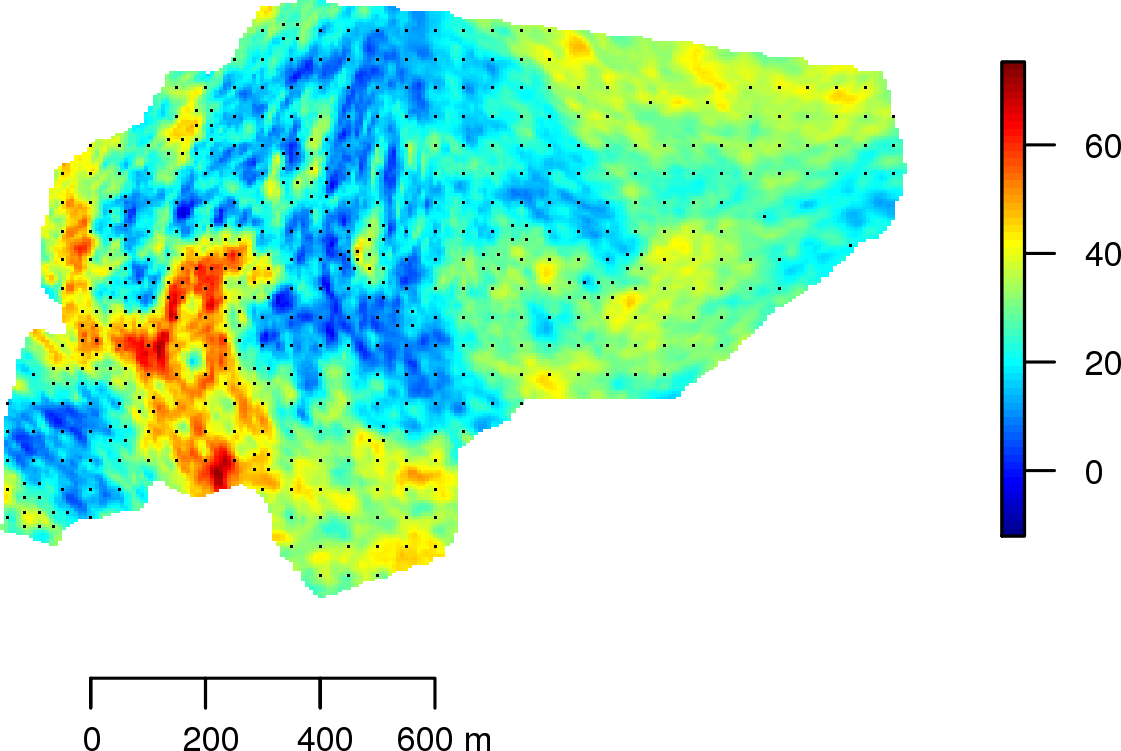}
                \caption{Simulation \#2}\label{Fig5b}
        \end{subfigure}
        
        \medskip
        
        \begin{subfigure}[h]{0.35\textwidth}
                \centering
                \includegraphics[width=1\textwidth]{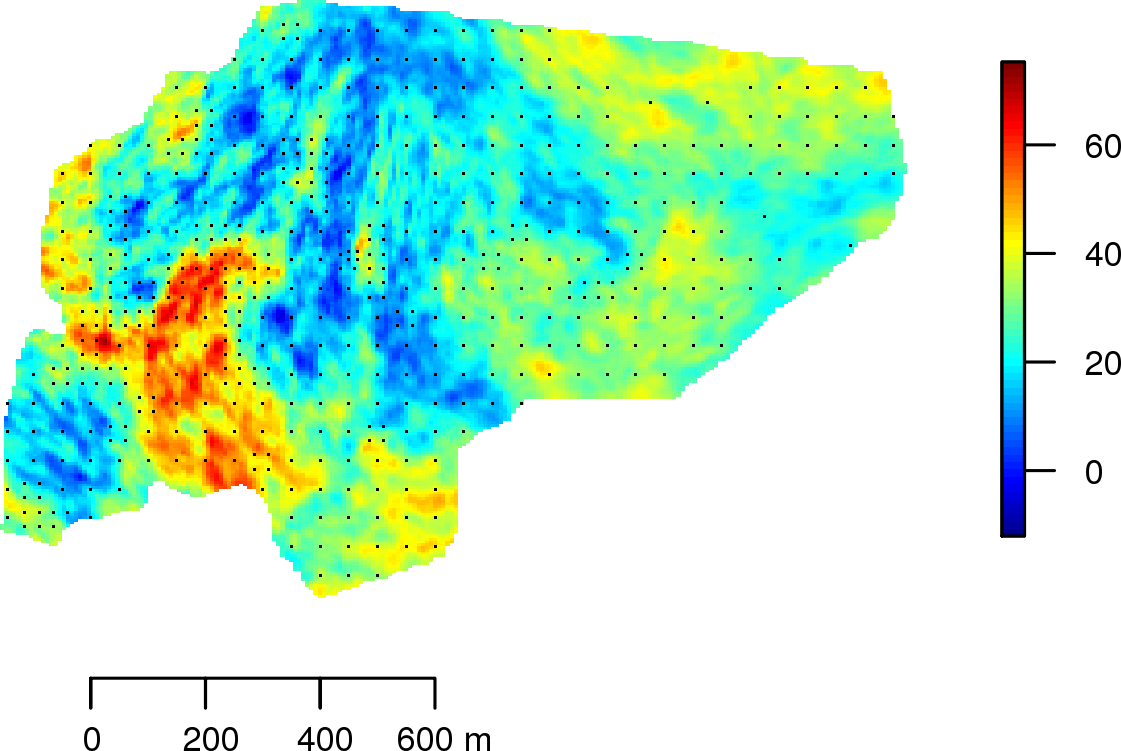}
                \caption{Simulation \#3}\label{Fig5c}
        \end{subfigure}
         \qquad      
        \begin{subfigure}[h]{0.35\textwidth}
                \centering
                \includegraphics[width=1\textwidth]{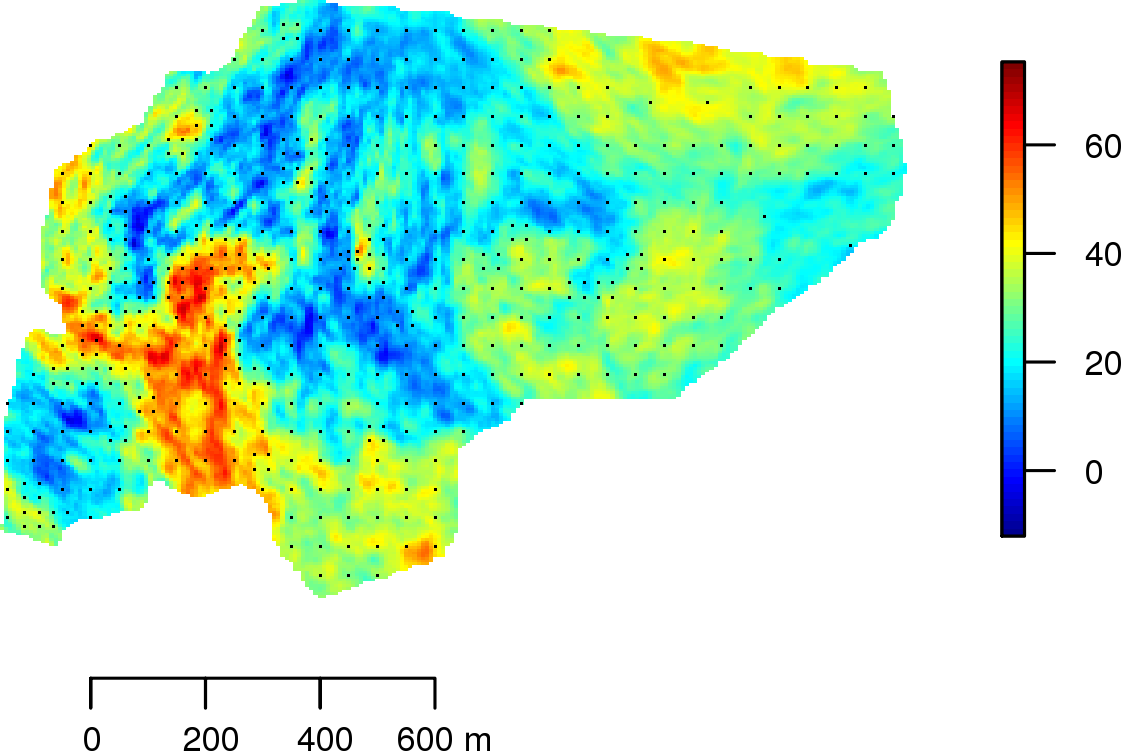}
                \caption{Simulation \#4}\label{Fig5d}
        \end{subfigure}
        
        \caption{Conditional simulations based on the estimated non-stationary model. (Potassium concentration data)}\label{Fig5}
\end{figure}

The kriged values and the kriging standard deviations for the stationary and non-stationary models are shown in Figure \ref{Fig4}. The overall look of the predicted values and prediction standard deviations associated with each model differ notably. In particular, the proposed method takes into account certain local characteristics of the regionalization that the stationary approach is unable to retrieve. This example shows the ability of our non-stationary approach to manage data with locally varying anisotropy. Figure \ref{Fig5} shows some conditional simulations in the Gaussian framework, based on the estimated non-stationary model.

\newpage
To assess the predictive performances of our approach, the regionalized variable is predicted at a hold-out sample (1000 points). Table \ref{Tab1} presents the summary statistics for the external validation results using the classical stationary approach and the new proposed one. Some well-known discrepancy measures are used, namely the Mean Absolute Error (MAE), the Root Mean Square Error (RMSE), the Normalized Mean Square Error (NMSE), the Logarithmic Score (LogS) and the Continued Rank Probability Score (CRPS). For RMSE, LogS and CRPS, the smaller the better; for MAE, the nearer to zero the better; for NMSE the nearer to one the better. Table \ref{Tab1} shows that the proposed approach outperforms the stationary one with respect to all the measures. The cost of non-using the non-stationary approach in this case can be substantial: in average the prediction  at validation locations is about 22\% better for the non-stationary model than for the stationary model, in terms of RMSE. 

\begin{table}[h!]
\begin{center}
\begin{tabular}{lcc}
\hline
         & Stationary Model& Non-Stationary Model  \\
         \hline

    Mean Absolute Error   & 2.86   & 2.12 \\
             
    Root Mean Square Error  & 3.88    & 3.17 \\
    
    Normalized Mean Square Error &0.19  & 0.30   \\
    
    Logarithmic Score & 6400   & 5268   \\
    
    Continued Rank Probability Score & 4.97   & 3.28  \\
    \hline
\end{tabular}
\end{center}
\caption{External validation  on a set of 1000 locations. (Potassium concentration data)}
\label{Tab1}
\end{table}

\newpage

\newpage
\subsection{Rainfall Data Example}
\label{ssec7}

We now consider the well known Swiss Rainfall data. The data corresponds to the record of rainfall (in mm) measured on 8th May 1986 at 467 fixed locations across Switzerland. We randomly partition the location of the data into a model calibration data subset of 400 values and a model validation data subset of 67 values.

\begin{figure}[h!]
        \centering
        \begin{subfigure}[h!]{0.35\textwidth}
                \centering
                \includegraphics[width=1\textwidth]{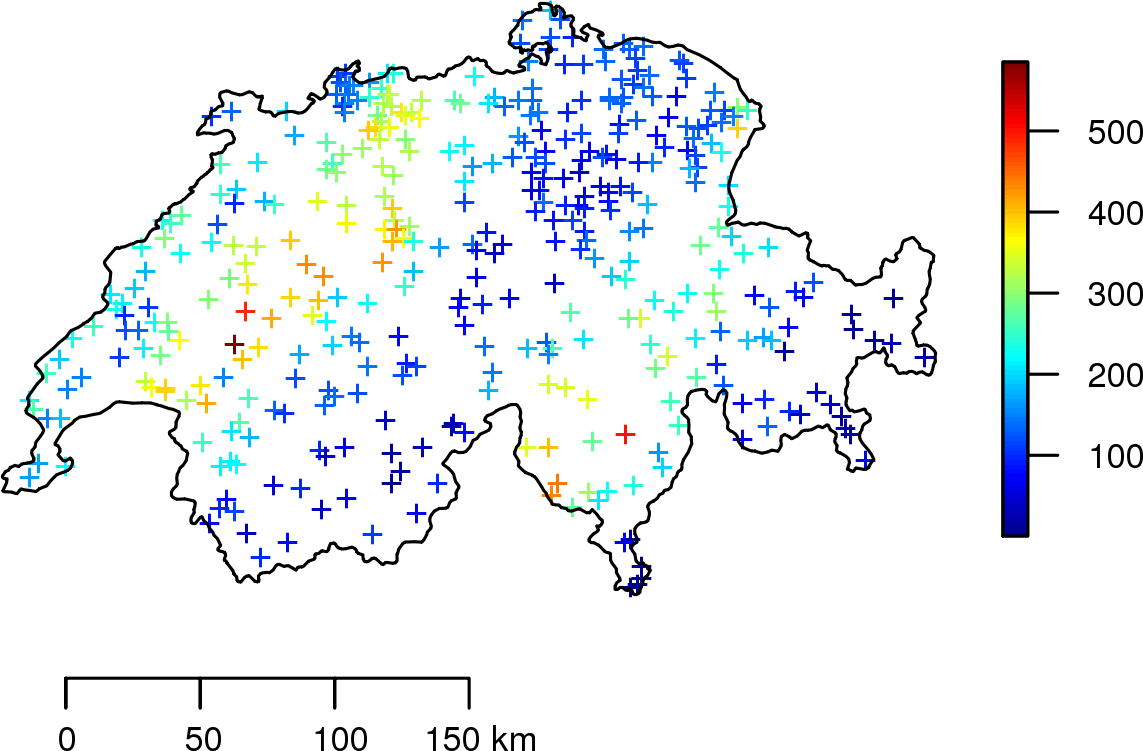}
                \caption{}\label{Fig6a}
        \end{subfigure}
         \qquad     
        \begin{subfigure}[h!]{0.35\textwidth}
                \centering
                \includegraphics[width=1\textwidth]{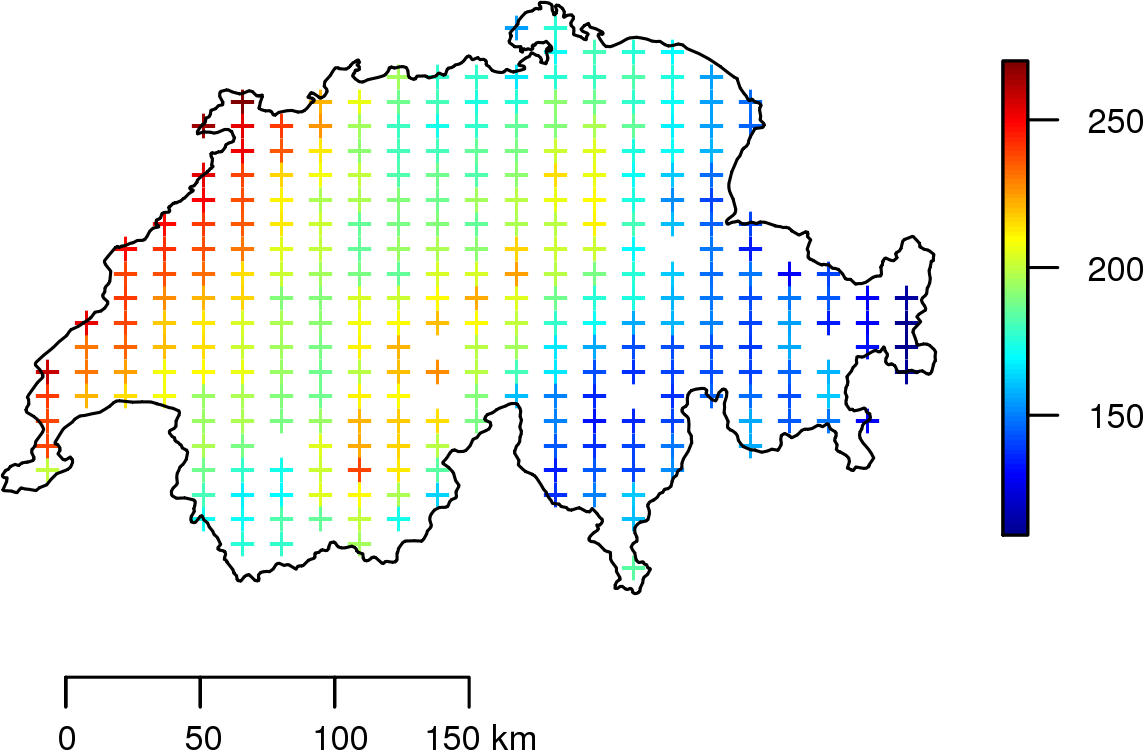}
                \caption{}\label{Fig6b}
        \end{subfigure}
        
        \begin{subfigure}[h!]{0.35\textwidth}
                \centering
                \includegraphics[width=1\textwidth]{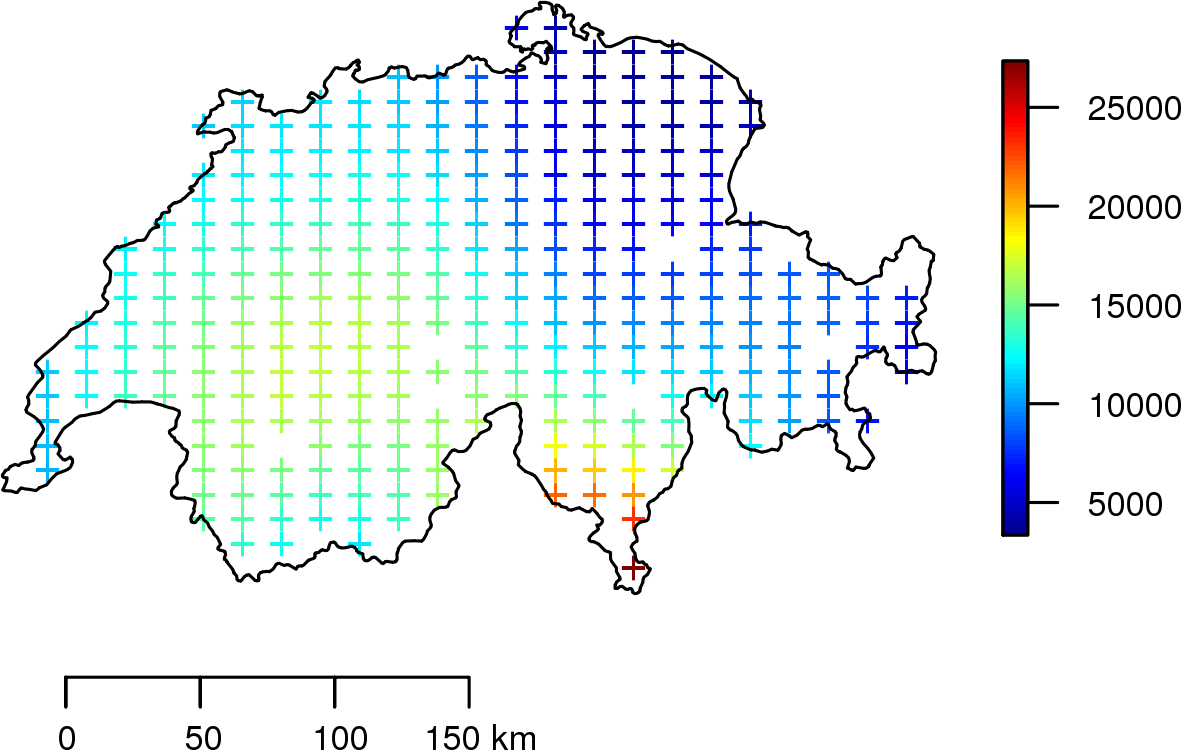}
                \caption{}\label{Fig6c}%
        \end{subfigure}
        \qquad 
        \begin{subfigure}[h!]{0.35\textwidth}
                \centering
                \includegraphics[width=1\textwidth]{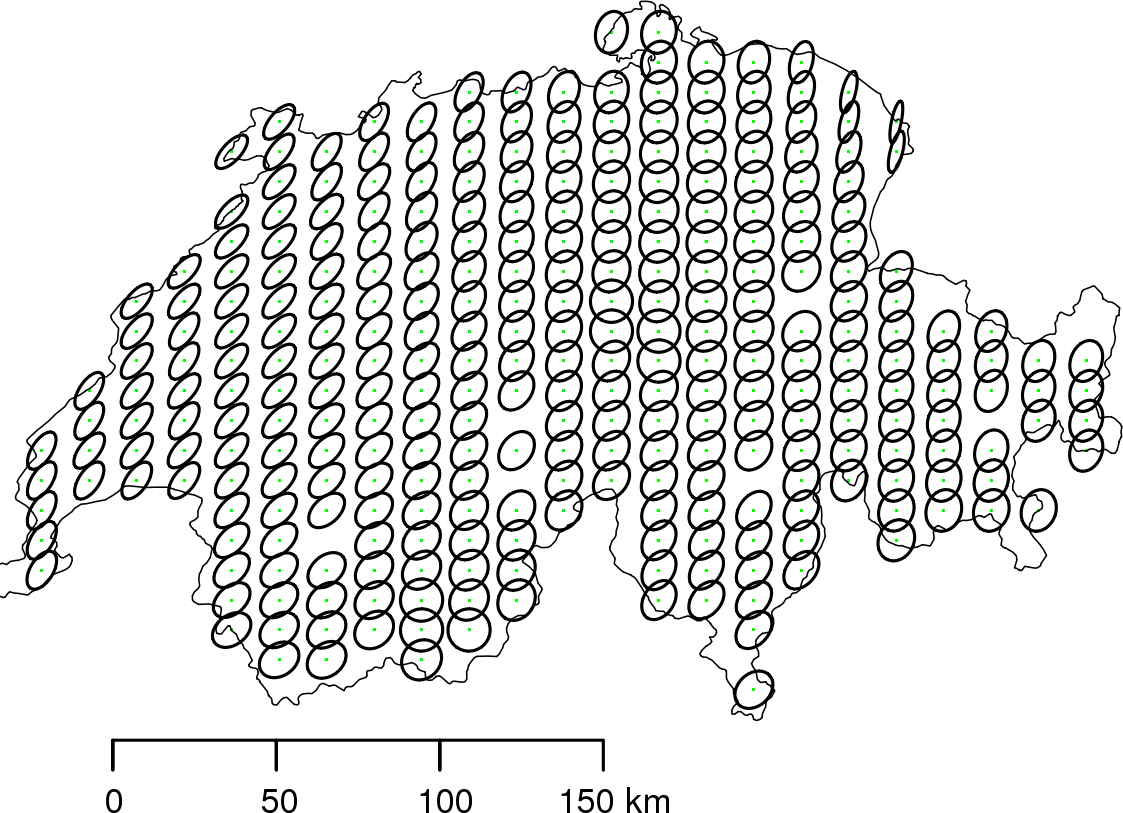}
                \caption{}\label{Fig6d}
        \end{subfigure}
        \caption{(a) Training data. (b) Estimated mean function $\widehat{\mbox{m}}(.)$ at anchor points. (c) Estimated variance function $\widehat{\sigma}^2(.)$ at anchor points. (d) Estimated anisotropy function $\widehat{\mathbf{\Sigma}}(.)$ at anchor points where the ellipses were scaled to ease vizualisation. (Rainfall data)}\label{Fig6}
\end{figure}

Figure \ref{Fig6a} shows the plot of the rainfall training data and suggests the presence of a global anisotropy direction in the data. This is confirmed when computing directional experimental variograms, in stationary framework. The orientation of maximum continuity for the global stationary model is along the direction South/West - North/East. \citet{Al06} have previously corroborated this conclusion. However, the proposed non-stationary method shows that we cannot consider the anisotropy to be globally constant. Figures \ref{Fig6b}, \ref{Fig6c} and \ref{Fig6d} show respectively the estimated parameters $\widehat{m}(.)$, $\widehat{\sigma}^2(.)$ and $\widehat{\mathbf{\Sigma}}(.)$ at anchor points. These raw estimates are  based on  the non-stationary exponential covariance function (Example \ref{Ex1}).  Figure \ref{Fig6d} reveals a locally varying geometric anisotropy including the main direction of continuity discovered by the stationnary approach. Note that global geometric anisotropy cannot describe local directional features of a spatial surface, only global ones.

\begin{figure}[h!]
        \centering
        \begin{subfigure}[h!]{0.35\textwidth}
                \centering
                \includegraphics[width=1\textwidth]{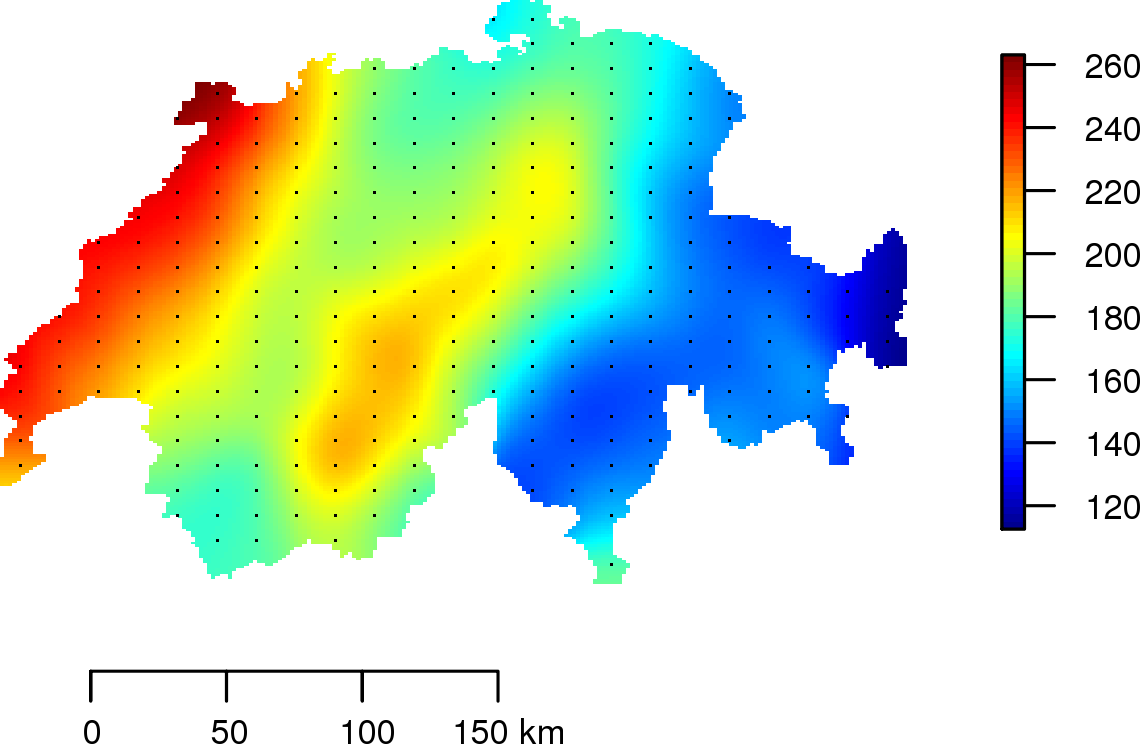}
                \caption{}\label{Fig7a}
        \end{subfigure}
         \quad     
        \begin{subfigure}[h!]{0.35\textwidth}
                \centering
                \includegraphics[width=1\textwidth]{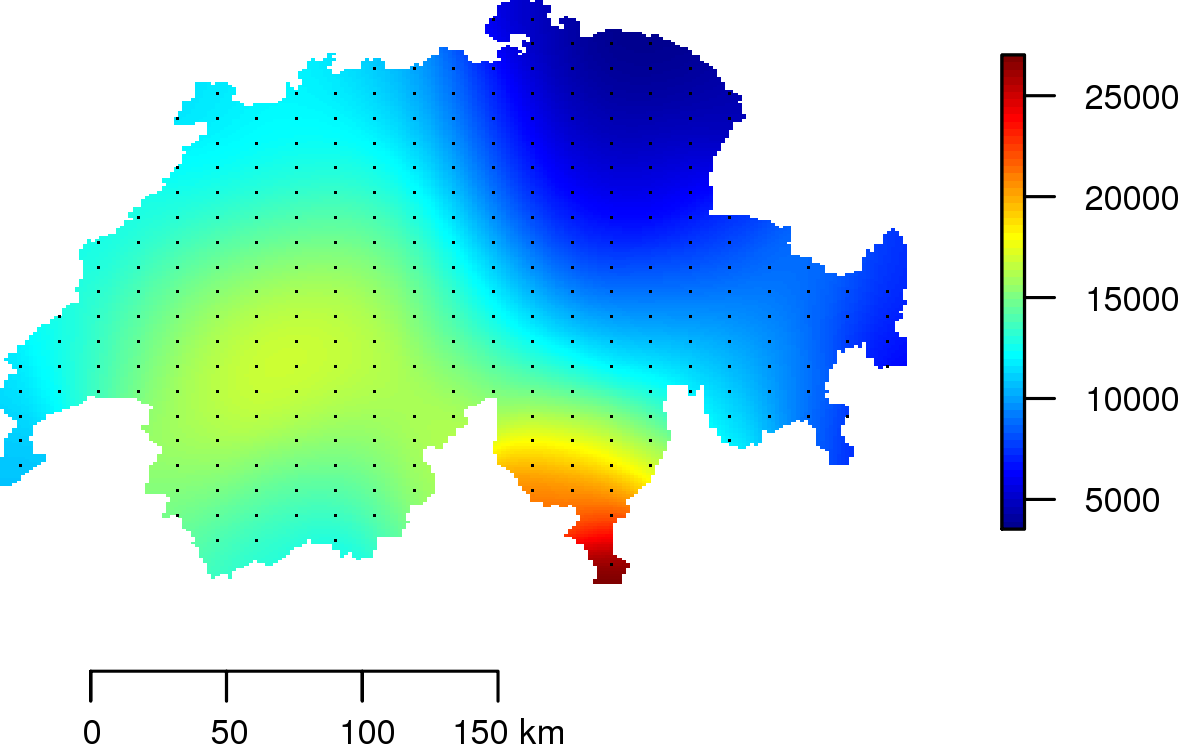}
                \caption{}\label{Fig7b}
        \end{subfigure}
        
        \begin{subfigure}[h!]{0.35\textwidth}
                \centering
                \includegraphics[width=1\textwidth]{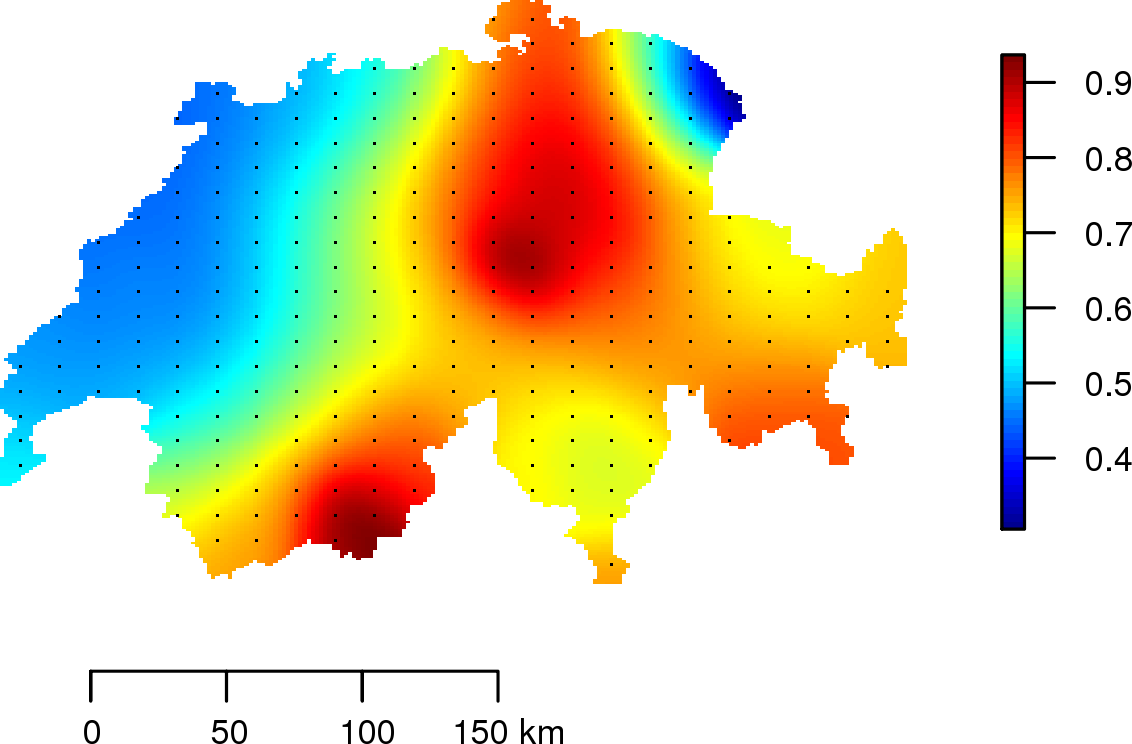}
                \caption{}\label{Fig7c}%
        \end{subfigure}
        \quad 
        \begin{subfigure}[h!]{0.35\textwidth}
                \centering
                \includegraphics[width=1\textwidth]{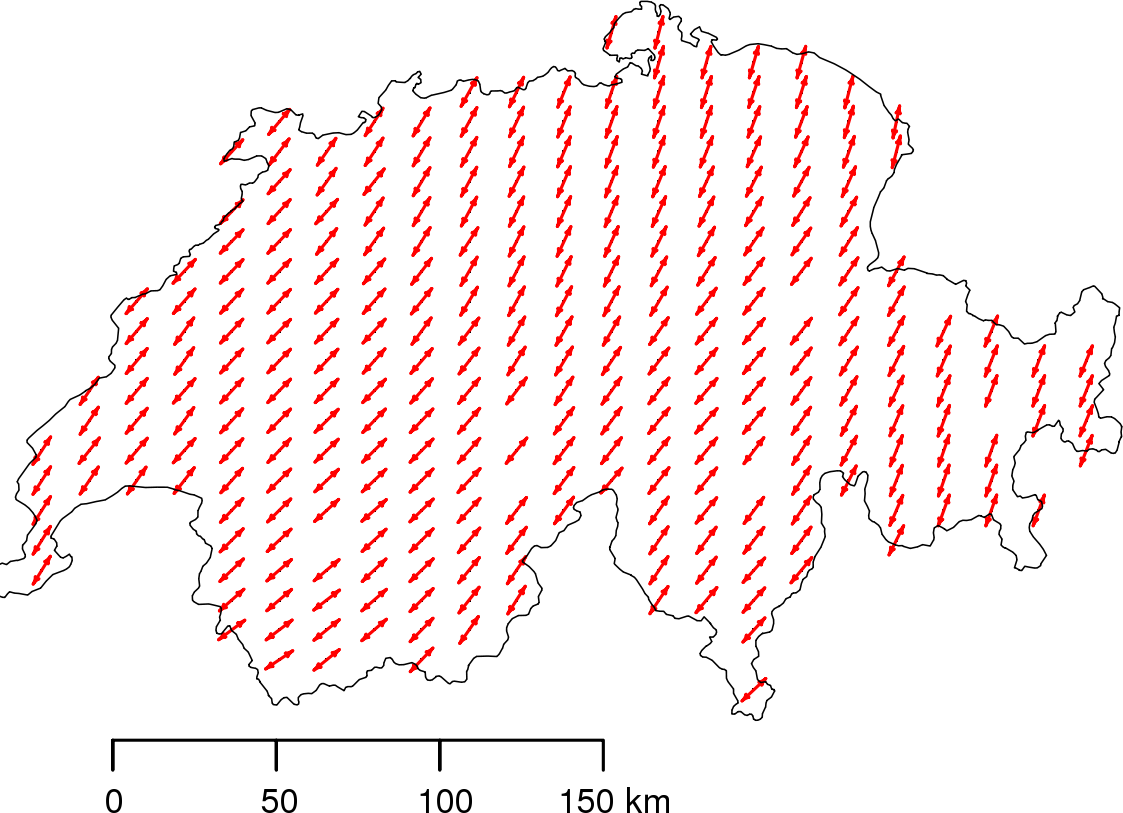}
                \caption{}\label{Fig7d}
        \end{subfigure}
        \caption{Rainfall data: Smoothed parameters over the domain of interest: (a) mean, (b) variance, (c) anisotropy ratio, (d) azimuth.}\label{Fig7}
\end{figure}

The selection procedure presented in Section \ref{ssec3} leads to take as bandwidth $\epsilon=46$ km. Figure \ref{Fig7} gives the maps of smoothed parameters over the entire domain of study: mean, variance, anisotropy ratio and azimuth. The optimal smoothing bandwidth associated to the Gaussian kernel smoothing corresponds to $\delta=11$ km, following the selection procedure described in Section \ref{ssec3}.

\begin{figure}[h!]
        \centering
        \begin{subfigure}[h]{0.35\textwidth}
                \centering
                \includegraphics[width=1\textwidth]{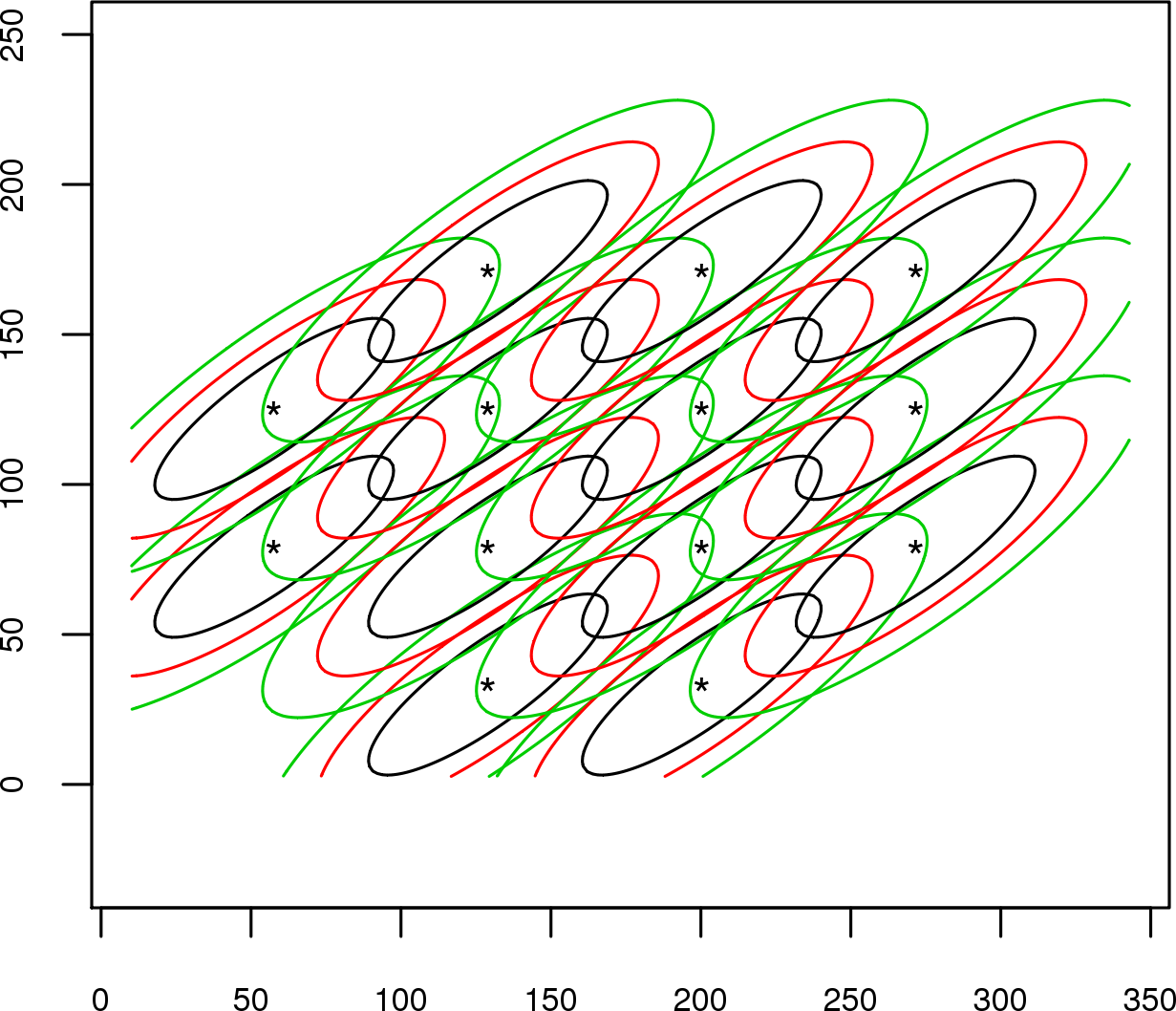}
                \caption{}\label{Fig8a}
        \end{subfigure}
        \qquad
        \begin{subfigure}[h]{0.35\textwidth}
                \centering
                {\includegraphics[width=1\textwidth]{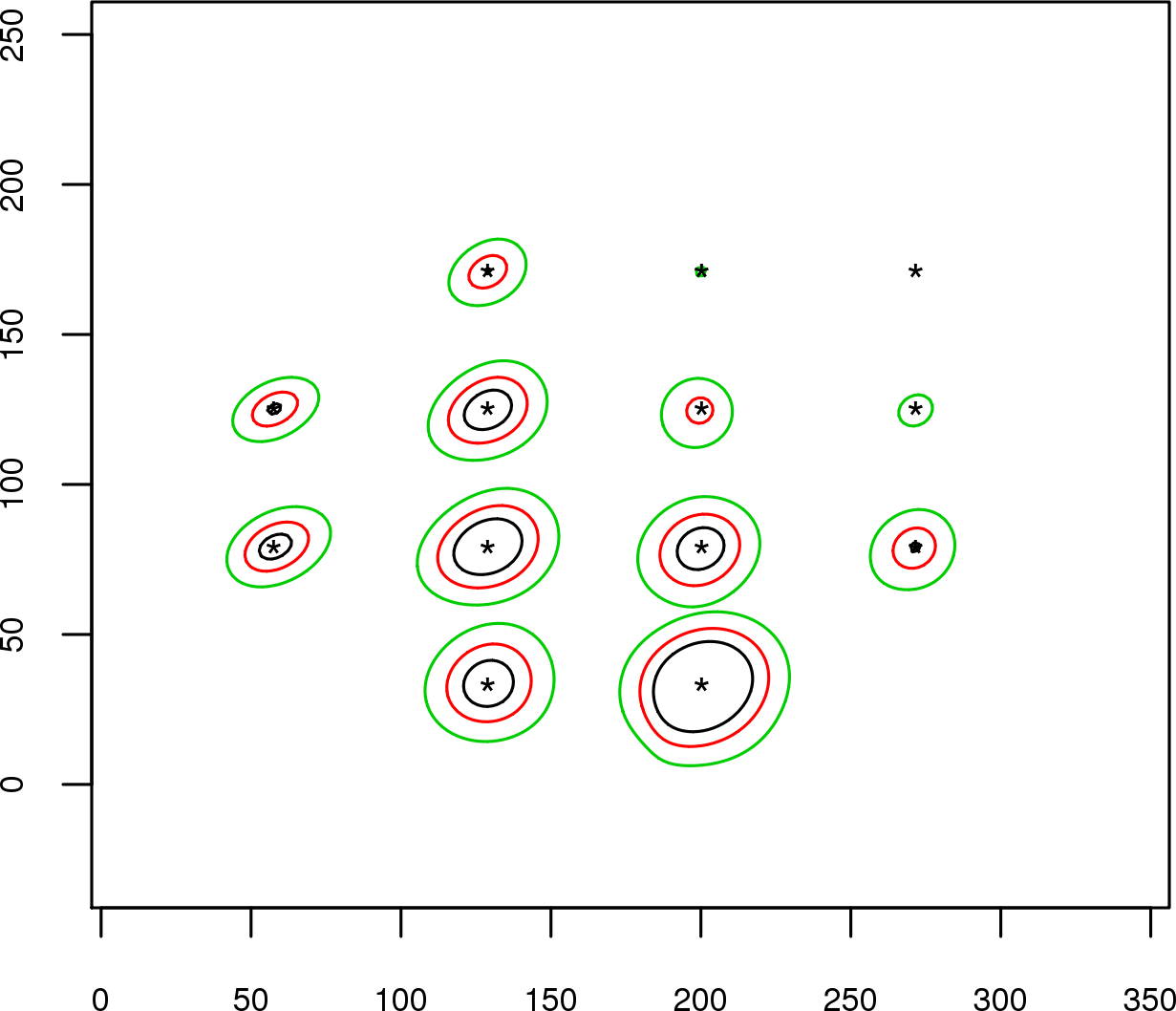}}
                \caption{}\label{Fig8b}
        \end{subfigure}
        \caption{Covariance  level contours at few points for the estimated stationary and non-stationary models (a, b). Level contours correspond to the values: $10000$ (black), $8000$ (red) and $6000$ (green). (Rainfall data)}\label{Fig8}
\end{figure}

Under stationarity and non-stationarity assumptions, a visualization of the estimated covariance structure at few locations is given by Figure \ref{Fig8}. The variation of the non-stationary covariance from one location to another is visible. The stationary model is a geometric anisotropic spherical model with direction of maximum continuity correspond to 27 degrees while the non-stationary model corresponds to the non-stationary exponential covariance (Example \ref{Ex1}).

\begin{figure}[h!]
        \centering
        \begin{subfigure}[h!]{0.35\textwidth}
                \centering
                \includegraphics[width=1\textwidth]{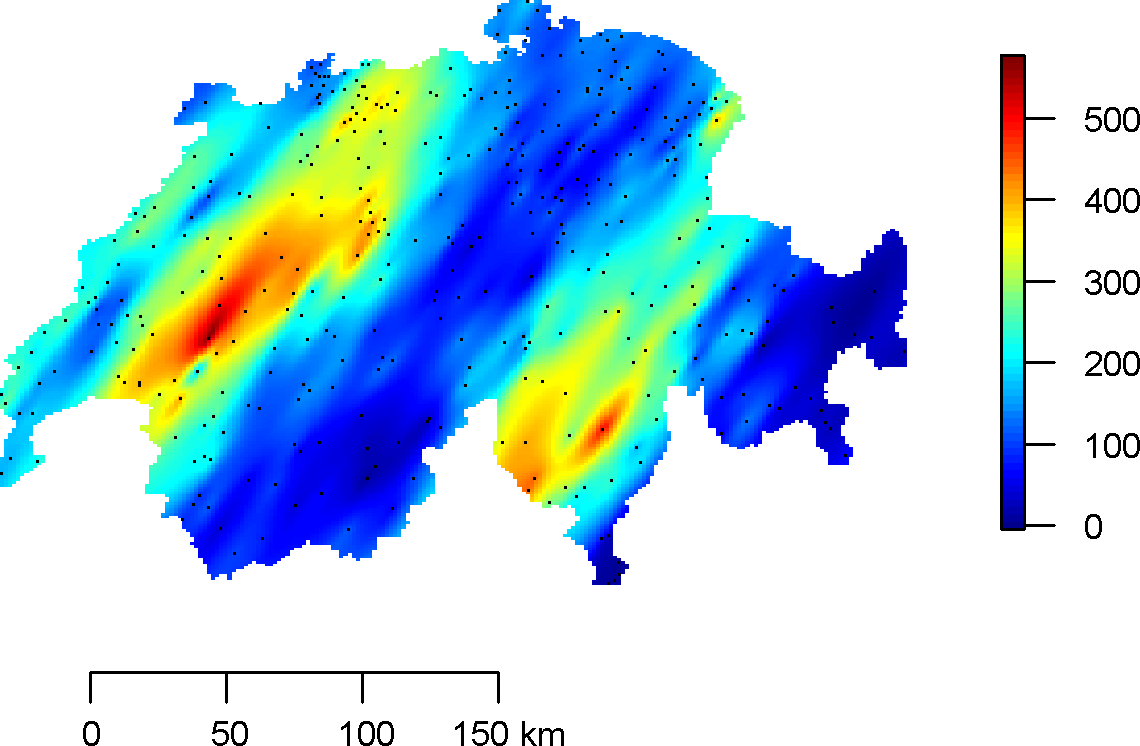}
                \caption{}\label{Fig9a}
        \end{subfigure}
        \qquad 
        \begin{subfigure}[h!]{0.35\textwidth}
                \centering
                \includegraphics[width=1\textwidth]{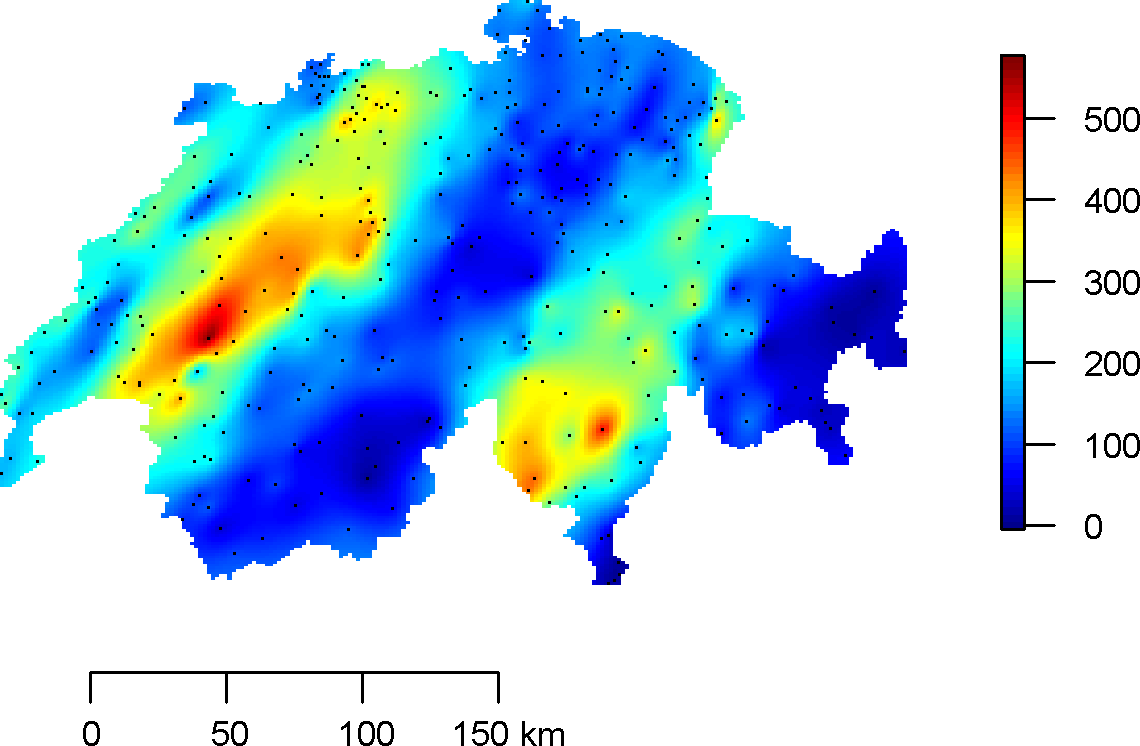}
                \caption{}\label{Fig9b}
        \end{subfigure}
        
        \medskip
        
        \begin{subfigure}[h!]{0.35\textwidth}
                \centering
                \includegraphics[width=1\textwidth]{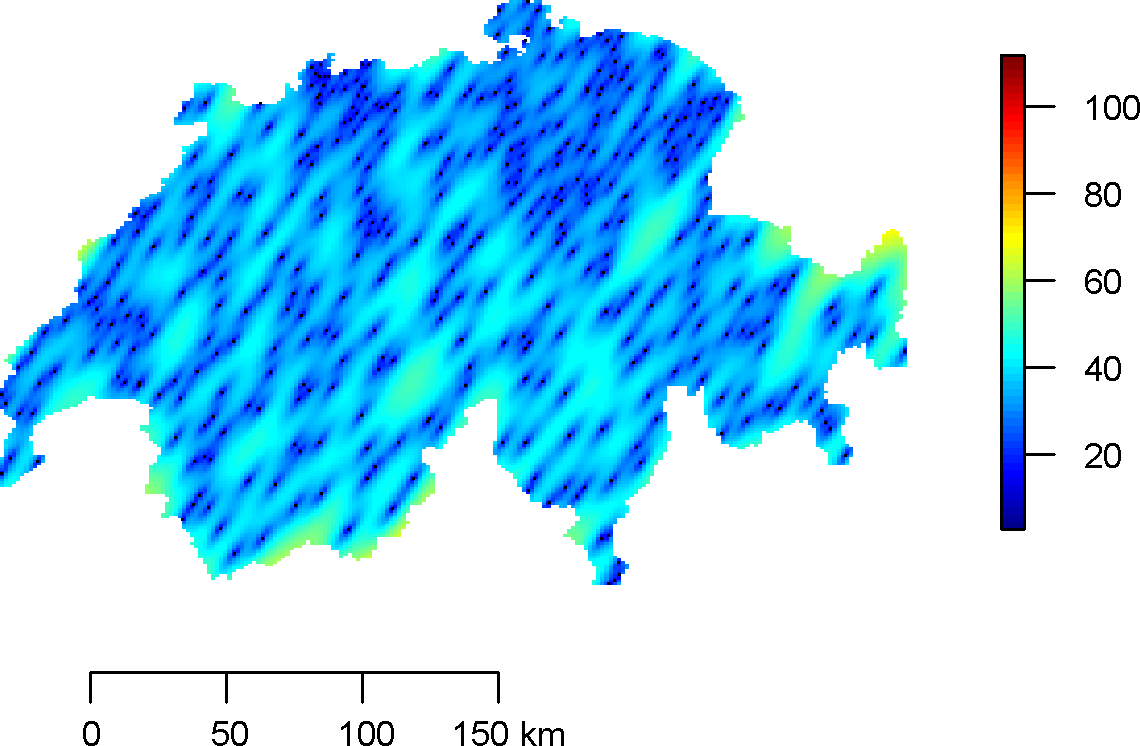}
                \caption{}\label{Fig9c}
        \end{subfigure}
        \qquad 
        \begin{subfigure}[h!]{0.35\textwidth}
                \centering
                \includegraphics[width=1\textwidth]{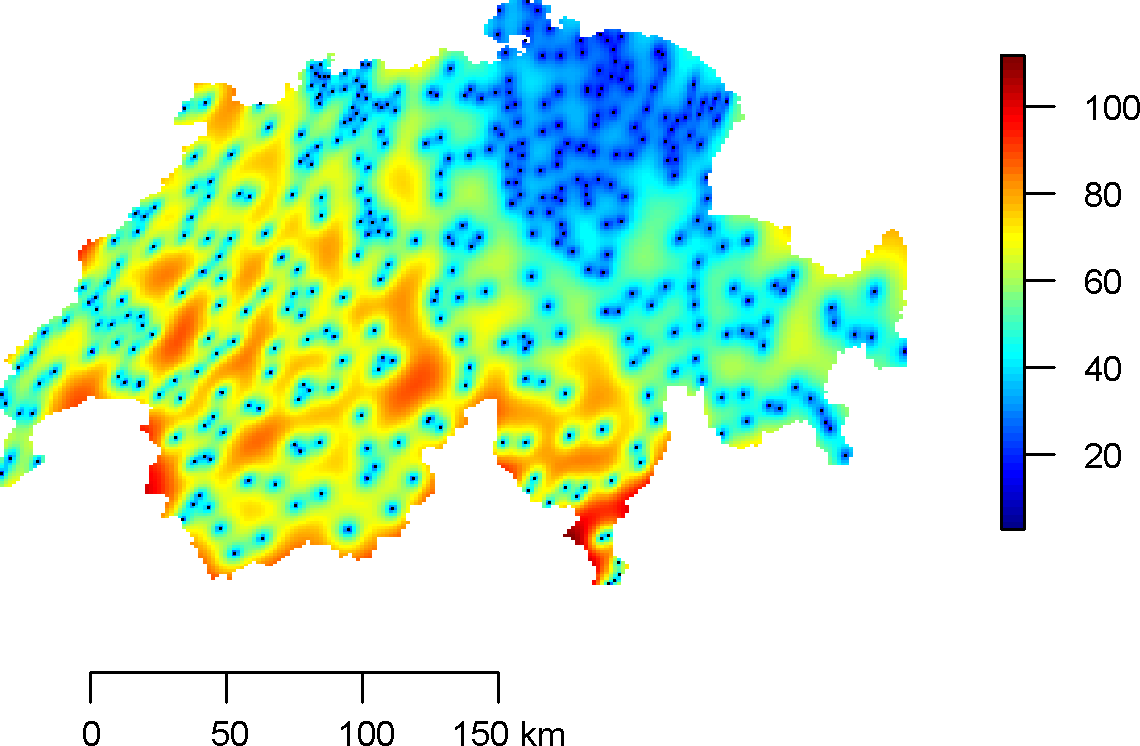}
                \caption{}\label{Fig9d}
        \end{subfigure}
        \caption{(a,b) Predictions and prediction standard deviations for the stationary model. (c,d) Predictions and prediction standard deviations for the non-stationary model. (Rainfall data)}\label{Fig9}
\end{figure}

\begin{figure}[h!]
        \centering      
        \begin{subfigure}[h]{0.35\textwidth}
                \centering
                \includegraphics[width=1\textwidth]{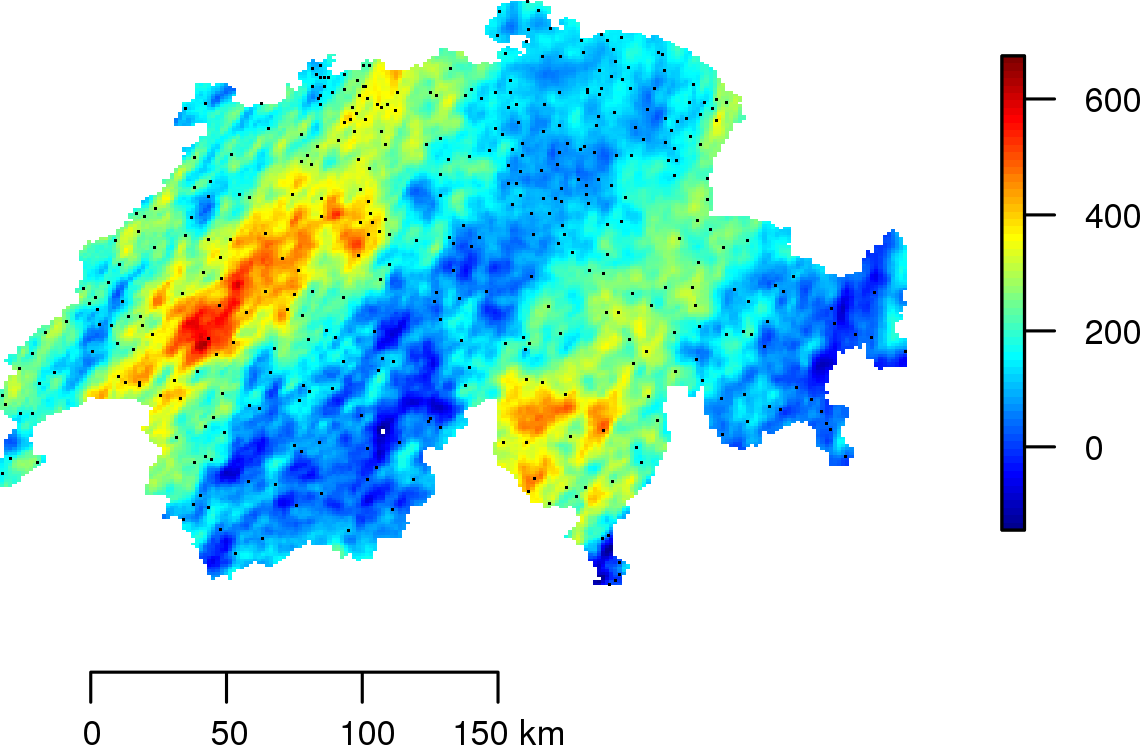}
                \caption{Simulation \#1}\label{Fig10a}
        \end{subfigure}
         \qquad    
        \begin{subfigure}[h]{0.35\textwidth}
                \centering
                \includegraphics[width=1\textwidth]{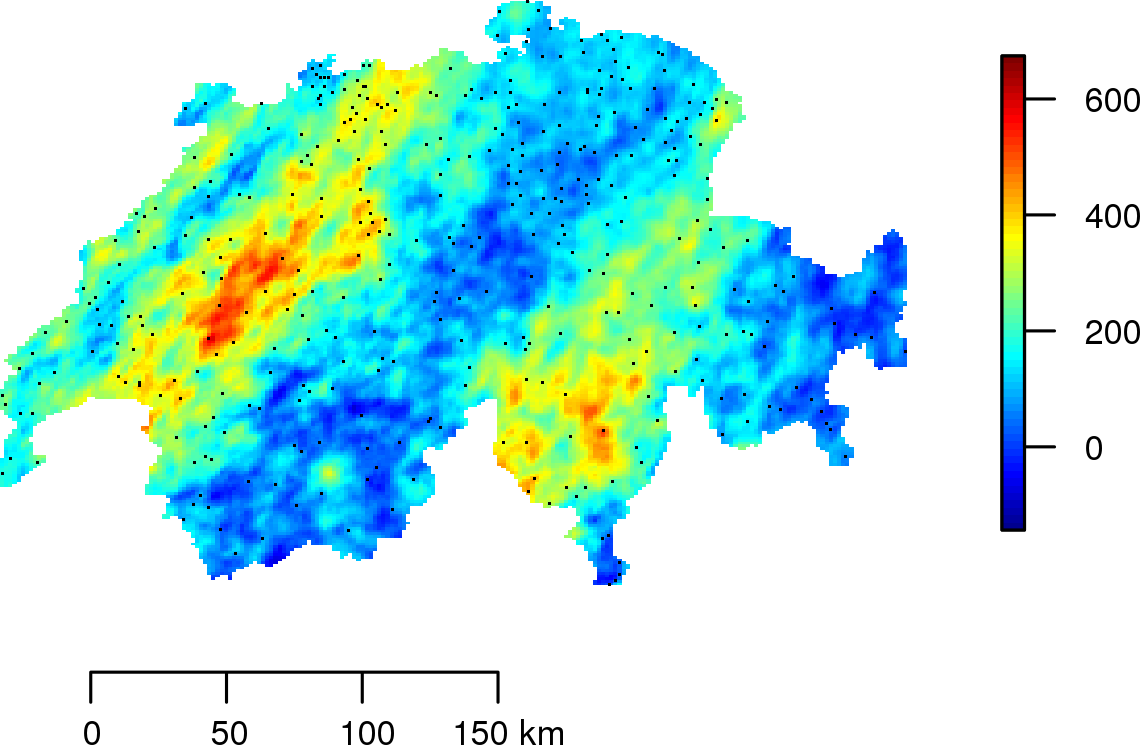}
                \caption{Simulation \#2}\label{Fig10b}
        \end{subfigure}
        
        \medskip
        
        \begin{subfigure}[h]{0.35\textwidth}
                \centering
                \includegraphics[width=1\textwidth]{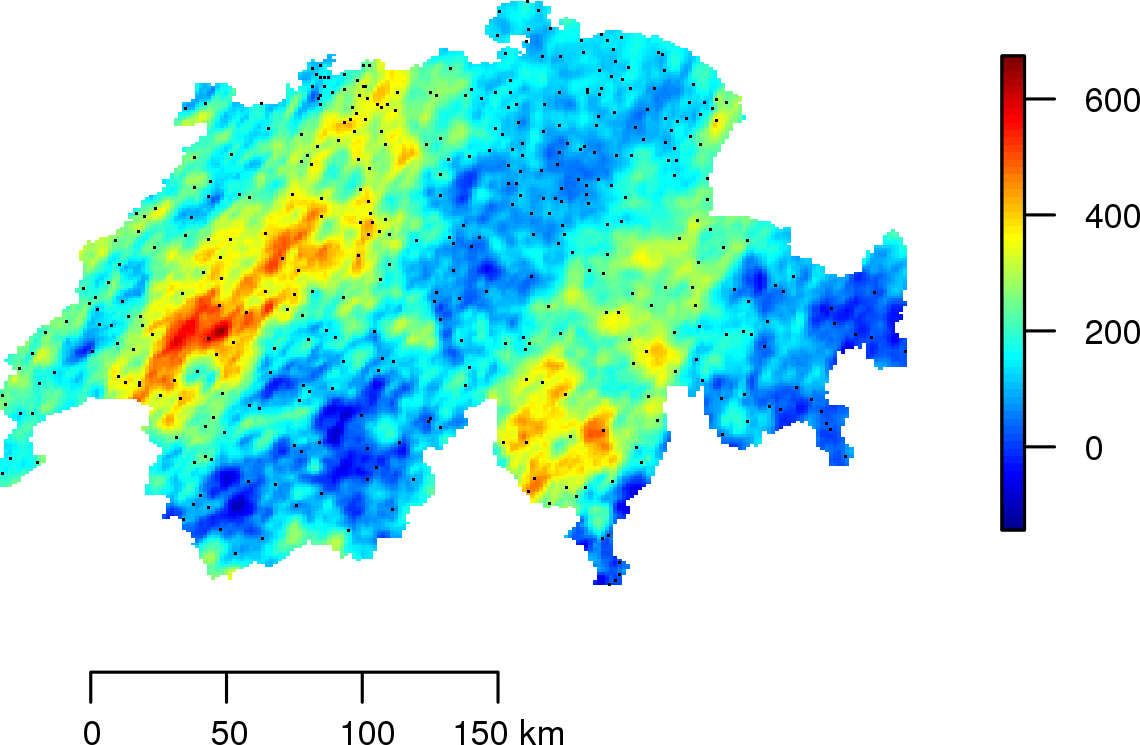}
                \caption{Simulation \#3}\label{Fig10c}
        \end{subfigure}
         \qquad      
        \begin{subfigure}[h]{0.35\textwidth}
                \centering
                \includegraphics[width=1\textwidth]{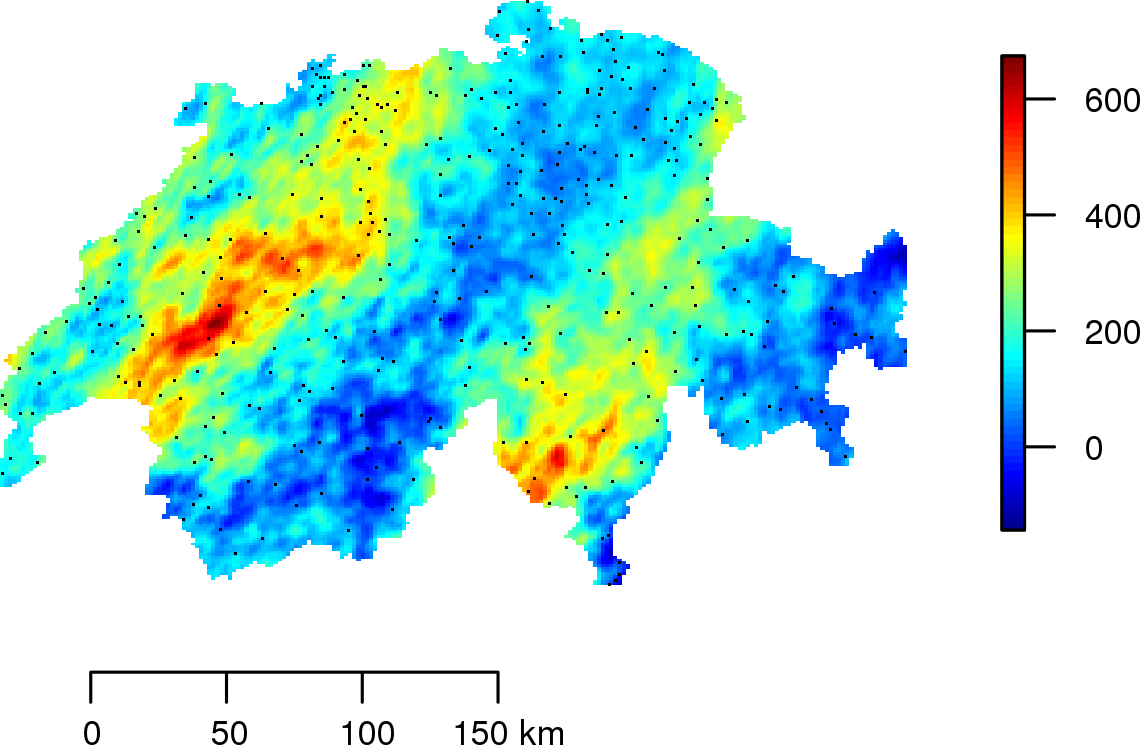}
                \caption{Simulation \#4}\label{Fig10d}
        \end{subfigure}
        
        \caption{Conditional simulations based on the estimated non-stationary model. (Rainfall data)}\label{Fig10}
\end{figure}

The predictions and prediction standard deviations for the estimated stationary, estimated non-stationary models are shown in figure \ref{Fig9}. The differences between kriging estimates and kriging standard deviation maps  are visibly marked. The stationary method provides an oversimplified version of the kriging standard deviations because it does not take into account variations of variability. Figure \ref{Fig10} shows some conditional simulations in the Gaussian context of the estimated non-stationary model.

\newpage
Table \ref{Tab2} summarizes the results for the predictive performance statistics computed on the validation data set (67 observations). The cost of not using a non-stationary model is not negligible. For example, the estimated stationary model is $13\%$ worse than the estimated non-stationary model in terms of Root Mean Square Error.

\begin{table}[h!]
\begin{center}
\begin{tabular}{lcc}
\hline
         & Stationary Model& Non-Stationary Model  \\
         \hline
    Mean Absolute Error   & 42.21   & 37.10 \\
            
    Root Mean Square Error  & 56.80    & 50.90 \\
    
    Normalized Mean Square Error &2.24  & 0.70   \\
    
    Logarithmic Score & 753  & 710   \\
    
    Continued Rank Probability Score & 53.31   & 50.60  \\
    \hline
\end{tabular}
\end{center}
\caption{External validation on a set of 67 locations. (Rainfall data)}
\label{Tab2}
\end{table}

\section{Concluding remarks}
\label{sec7}

The proposed convolution approach  provides in terms of modelling a new model for second order non-stationary Random Functions. This new model generalizes the classical convolution model and provides explicit and flexible classes of non-stationary covariance functions known from the literature. Moreover, a statistical methodology for estimating these latter is developed. The estimation method offers an integrated treatment of all aspects of non-stationarity (mean, variance, spatial continuity) in the modelling process. The estimation procedure relies on the mild hypothesis of quasi-stationarity and does not impose any distributional assumptions except the existence of the first and second moment. Furthermore, no matrix inversions or determinants are required, and hence the  inference is practicable even for large datasets. The proposed method provides an exploratory analysis tool for the non-stationarity. Indeed, mapping of the parameter values (mean, variance, azimuth) by location allows to exhibit or describe the non-stationarity. A plot of variance  versus mean allows to identify the common relationships between the mean and variance which is known as proportional effect \citep{Mat71}.

The performance of our proposed method has been demonstrated on two real datasets: soil and rainfall data. The two applications have revealed an increased prediction accuracy when compared to the standard stationary method, and demonstrated the ability to extract the underlying non-stationarity from a single realization. A comparison of predictions and prediction standard deviations maps indicates that our non-stationary method captures some varying spatial features (such as locally varying anisotropy) in the data that can be not present detected with the stationary method, the resulting outcome appears much more realistic.  Beyond the spatial predictions, we also show how conditional simulations can be carried out in this non-stationary framework.

In the proposed convolution method, the estimation relies on the  local variogram non-parametric kernel estimator. To better adapt to the variable sampling density in the domain of observations, it would be interesting to work with a non-parametric locally adaptive kernel estimator. The idea is to increase the bandwidth in low sample density areas and to narrow it in highly sampled areas. The local stationarity assumption is the basis of the proposed methodology, then it works well for smoothly varying non-stationarity. However it can be difficult to apply on sparse data or data that present rapid or abrupt spatial structure variations. In these cases, it may be advisable to proceed under a stationary framework. It would be interesting to extend the proposed method for inclusion of covariates. This can be achieved by setting the variance and anisotropy functions for covariates. This can be done by following the work of \citet{Net13} and \citet{Ris14}. To date, the non-stationary modelling convolution approach does not provide closed-form non-stationary covariance functions with compact support, this remains an open problem. Indeed, the use of such covariances considerably reduce the computational burden of kriging when dealing with large data sets.

\section*{Acknowledgements}
The authors would like to thank Dr Budiman Minasny at the Faculty of Agriculture \& Environment at the University of Sydney in Australia, for providing the first data used in this paper.

\appendix
\section{}
\label{appendix-sec1}

\subsubsection*{Proof of Proposition \ref{Prop1}}

For all $\mathbf{x} \in G$, the expectation of $Z$ is
\begin{eqnarray*} \label{EqA3}
\mathds{E}(Z(\mathbf{x}))&=& \mathds{E} \left(\int_{\mathds{R}^p}f_{\mathbf{x}}(\mathbf{u};T(\mathbf{u}))W(d\mathbf{u})\right) \\
              &=& \int_{\mathds{R}^p}\mathds{E}\left( f_{\mathbf{x}}(\mathbf{u};T(\mathbf{u}))W(d\mathbf{u})\right) \  \mbox{(Fubini)} \\
              &=& \int_{\mathds{R}^p}\mathds{E}\left( f_\mathbf{x}(\mathbf{u};T(\mathbf{u}))\right)\mathds{E}\left( W(d\mathbf{u})\right), \  \mbox{since  } \ T(\mathbf{u})\  \mbox{and  }  \ W(d\mathbf{u})  \ \mbox{are independent  }\\
              &=& 0, \ \mbox{car} \quad \mathds{E}\left( W(d\mathbf{u})\right)=0 \  \mbox{and} \  \mathds{E}\left( f_\mathbf{x}(\mathbf{u};T(\mathbf{u}))\right) \mbox{is assumed finite and integrable on} \ \mathds{R}^p.
\end{eqnarray*}

$Z$ being centered, for all $\mathbf{x},\mathbf{y} \in G$, its covariance is
\begin{eqnarray*}\label{EqA4}
\mathds{C}\mbox{ov}(Z(\mathbf{x}),Z(\mathbf{y}))&=&\mathds{E}(Z(\mathbf{x})Z(\mathbf{y}))\\
               &=& \int_{\mathds{R}^p}\int_{\mathds{R}^p}\mathds{E}(f_{\mathbf{x}}(\mathbf{u};T(\mathbf{u}))f_{\mathbf{y}}(\mathbf{v};T(\mathbf{v}))W(d\mathbf{u})W(d\mathbf{v})) \ \mbox{(Fubini)}\\
              &=& \int_{\mathds{R}^p}\int_{\mathds{R}^p}\mathds{E}(f_{\mathbf{x}}(\mathbf{u};T(\mathbf{u}))f_{\mathbf{y}}(\mathbf{v};T(\mathbf{v})))\mathds{E}(W(d\mathbf{u})W(d\mathbf{v})), \  \mbox{since  } \ T(\mathbf{u})\  \mbox{and  }  \ W(d\mathbf{u})  \ \mbox{are independent  } \\
              &=& \lambda \int_{\mathds{R}^p}\int_{\mathds{R}^p}\mathds{E}(f_{\mathbf{x}}(\mathbf{u};T(\mathbf{u}))f_{\mathbf{y}}(\mathbf{v};T(\mathbf{v}))) \delta_{\mathbf{u}}(d\mathbf{v}) d\mathbf{u}\\
              &=& \int_{\mathcal{T}}\int_{\mathds{R}^p}f_{\mathbf{x}}(\mathbf{u};t)f_{\mathbf{y}}(\mathbf{u};t)d\mathbf{\mathbf{u}}M(dt)\\
              &=&  \int_{\mathcal{T}}C(\mathbf{x},\mathbf{y};t)M(dt).
\end{eqnarray*}

The integral $C(\mathbf{x},\mathbf{y};t)$ for all $t \in \mathcal{T}$ exists and is finite, because the product of two square integrable functions is integrable. Furthermore, it is easy to show that for all  $t \in \mathcal{T}$, $C(. , .;t)$ is positive definite on $\mathds{R}^p$. Since $\mathds{V}\left(Z(\mathbf{x})\right)<+\infty,  \ \forall \mathbf{x} \in G $, $C(. , .;t)$ is integrable on $\mathcal{T}$ with respect to the positive measure $M(.)$ for every pair $(\mathbf{x},\mathbf{y})\in \mathds{R}^p \times \mathds{R}^p $. Then,  $C: \mathds{R}^p \times \mathds{R}^p \rightarrow \mathds{R}, (\mathbf{x},\mathbf{y}) \mapsto C(\mathbf{x},\mathbf{y})=\int_{\mathcal{T}}C(\mathbf{x},\mathbf{y};t)M(dt)$ is a valid covariance function on $\mathds{R}^p \times \mathds{R}^p$, according to \citet{Mat86}.

The proof of the proposition \ref{Prop2} relies on the following lemma.

\begin{lemma}\label{lem1}
\begin{equation*}\label{EqA8}
\forall (\mathbf{x},\mathbf{y}) \in \mathds{R}^p \times \mathds{R}^p, \ \int_{\mathds{R}^p}k_\mathbf{x}(\mathbf{u};t)k_\mathbf{y}(\mathbf{u};t)d\mathbf{u}= \pi^{-\frac{p}{2}}t^{-p}{\left| \frac{\mathbf{\Sigma}_\mathbf{x}+ \mathbf{\Sigma}_\mathbf{y}}{2} \right|}^{-\frac{1}{2}}\exp\left(-\frac{Q_{\mathbf{xy}}(\mathbf{x}-\mathbf{y})}{t^2}\right).
\end{equation*}
\end{lemma}

\newpage
\textit{Proof}

Let $\tilde{C}(\mathbf{x},\mathbf{y};t)=\int_{\mathds{R}^p}k_\mathbf{x}(\mathbf{u};t)k_\mathbf{y}(\mathbf{u};t)d\mathbf{u}$.

Since $k_{\mathbf{x}}(.;t)$ is the density of the distribution $\mathcal{N}\left(\mathbf{x},\frac{t^2\mathbf{\Sigma}_\mathbf{x}}{4}\right)$, we have
\begin{equation*}\label{EqA6}
k_{\mathbf{x}}(\mathbf{u};t)=(2\pi)^{-\frac{p}{2}} {\left| \frac{t^2\mathbf{\Sigma}_{\mathbf{x}}}{4}\right|}^{-\frac{1}{2}} \exp\left(-\frac{1}{2}(\mathbf{u}-\mathbf{x})^{T}\left(\frac{t^2\mathbf{\Sigma}_{\mathbf{x}}}{4}\right)^{-1}(\mathbf{u}-\mathbf{x})\right).
\end{equation*}

Let $\mathbf{X} \sim \mathcal{N}\left(\mathbf{0},\frac{t^2\mathbf{\Sigma}_\mathbf{x}}{4}\right)$ and $\mathbf{Y} \sim \mathcal{N}\left(\mathbf{y},\frac{t^2\mathbf{\Sigma}_\mathbf{y}}{4}\right)$ two independent Gaussian random vectors with respective density functions $f_\mathbf{X}(.)$ and $f_\mathbf{Y}(.)$. Thus, we have: $f_\mathbf{X}(\mathbf{u}-\mathbf{x})=k_\mathbf{x}(\mathbf{u};t)$ and $f_\mathbf{Y}(\mathbf{u})=k_\mathbf{y}(\mathbf{u};t)$. Hence,
\begin{equation*}\label{EqC9}
\tilde{C}(\mathbf{x},\mathbf{y};t)=\int_{\mathds{R}^p}f_\mathbf{X}(\mathbf{x}-\mathbf{u})f_\mathbf{Y}(\mathbf{u})d\mathbf{u}=\int_{\mathds{R}^p}f_{\mathbf{X},\mathbf{Y}}(\mathbf{x}-\mathbf{u},\mathbf{u})d\mathbf{u}.
\end{equation*}

Now consider the $C^1$-diffeomorphism $(\mathbf{X},\mathbf{Y}) \rightarrow (\mathbf{U},\mathbf{V})\quad \mbox{with} \quad \mathbf{U}=\mathbf{X}+\mathbf{Y} \  \mbox{and} \ \mathbf{V}=\mathbf{Y}$, which has Jacobian $1$. We have
\begin{eqnarray*}\label{EqC10}
\int_{\mathds{R}^p}f_{\mathbf{X},\mathbf{Y}}(\mathbf{x}-\mathbf{u},\mathbf{u})d\mathbf{u}=\int_{\mathds{R}^p}f_{\mathbf{U},\mathbf{V}}((\mathbf{x}-\mathbf{u})+\mathbf{u},\mathbf{u})d\mathbf{u}=\int_{\mathds{R}^p}f_{\mathbf{U},\mathbf{V}}(\mathbf{x},\mathbf{u})d\mathbf{u}=f_\mathbf{U}(\mathbf{x})=f_{\mathbf{X}+\mathbf{Y}}(\mathbf{x}).
\end{eqnarray*}

Since $\mathbf{X} \sim \mathcal{N}\left(\mathbf{0},\frac{t^2\mathbf{\Sigma}_\mathbf{x}}{4}\right)$ and $\mathbf{Y} \sim \mathcal{N}\left(\mathbf{y},\frac{t^2\mathbf{\Sigma}_\mathbf{y}}{4}\right)$ are independent, then $\mathbf{X}+\mathbf{Y} \sim \mathcal{N}\left(\mathbf{y},\frac{t^2\mathbf{\Sigma}_\mathbf{x}}{4}+\frac{t^2\mathbf{\Sigma}_\mathbf{y}}{4}\right)$. Therefore,
\begin{eqnarray*}\label{EqC11}
\tilde{C}(\mathbf{x},\mathbf{y};t)&=&(2\pi)^{-\frac{p}{2}} {\left| \frac{t^2\mathbf{\Sigma}_\mathbf{x}+t^2\mathbf{\Sigma}_\mathbf{y}}{4} \right|}^{-\frac{1}{2}} \exp\left(-\frac{1}{2}(\mathbf{x}-\mathbf{y})^{T}\left(\frac{t^2\mathbf{\Sigma}_\mathbf{x}+t^2\mathbf{\Sigma}_\mathbf{y}}{4}\right)^{-1}(\mathbf{x}-\mathbf{y})\right)\\
&=& (2\pi)^{-\frac{p}{2}}t^{-p}{\left| \frac{\mathbf{\Sigma}_\mathbf{x}+ \mathbf{\Sigma}_\mathbf{y}}{2} \right|}^{-\frac{1}{2}}2^{\frac{p}{2}}\exp\left(-(\mathbf{x}-\mathbf{y})^{T}\left(\frac{\mathbf{\Sigma}_\mathbf{x}+\mathbf{\Sigma}_\mathbf{y}}{2}\right)^{-1}(\mathbf{x}-\mathbf{y})\frac{1}{t^2}\right)\\
&=& \pi^{-\frac{p}{2}}t^{-p}{\left| \frac{\mathbf{\Sigma}_\mathbf{x}+ \mathbf{\Sigma}_\mathbf{y}}{2} \right|}^{-\frac{1}{2}}\exp\left(-\frac{Q_{\mathbf{xy}}(\mathbf{x}-\mathbf{y})}{t^2}\right).
\end{eqnarray*}

\subsubsection*{Proof of proposition \ref{Prop2}}

The covariance $C(\mathbf{x},\mathbf{y};t)$ is given by
\begin{eqnarray*}\label{EqC7}
C(\mathbf{x},\mathbf{y};t)&=&\int_{\mathds{R}^p}f_\mathbf{x}(\mathbf{u};t)f_\mathbf{y}(\mathbf{u};t)d\mathbf{u}
        = g(\mathbf{x};t)g(\mathbf{y};t)\int_{\mathds{R}^p}k_\mathbf{x}(\mathbf{u};t)k_\mathbf{y}(\mathbf{u};t)d\mathbf{u}.
\end{eqnarray*}

Applying the lemma \ref{lem1}, we obtain
$$C(\mathbf{x},\mathbf{y};t)= g(\mathbf{x};t)g(\mathbf{y};t)\pi^{-\frac{p}{2}}t^{-p}{\left| \frac{\mathbf{\Sigma}_\mathbf{x}+ \mathbf{\Sigma}_\mathbf{y}}{2} \right|}^{-\frac{1}{2}}\exp\left(-\frac{Q_{\mathbf{xy}}(\mathbf{x}-\mathbf{y})}{t^2}\right).$$ 

Hence,
\begin{eqnarray*}
C^{NS}(\mathbf{x},\mathbf{y})&=&\int_{0}^{+\infty}C(\mathbf{x},\mathbf{y};t)M(dt)\\
&=& \pi^{-\frac{p}{2}}{\left| \frac{\mathbf{\Sigma}_\mathbf{x}+ \mathbf{\Sigma}_\mathbf{y}}{2} \right|}^{-\frac{1}{2}}\int_{0}^{+\infty}g(\mathbf{x};t)g(\mathbf{y};t)t^{-p}\exp\left(-\frac{Q_{\mathbf{xy}}(\mathbf{x}-\mathbf{y})}{t^2}\right)M(dt).
\end{eqnarray*}

\subsubsection*{Proof of Corollary \ref{Cor1}}

If $g(\mathbf{x};t) \propto 1 $ and $M(dt)=t^p\xi(dt)$, for $\xi(.)$ a finite positive measure on $\mathds{R}_{+}$ such that
$\displaystyle{R^{S}(\tau)=\int_0^{+\infty}\exp\left(-\frac{\tau^2}{t^2}\right)\xi(dt), \tau \geq 0}$,
then
\begin{eqnarray*}\label{EqA12}
C^{NS}(\mathbf{x},\mathbf{y})&\propto& \pi^{-\frac{p}{2}}{\left| \frac{\Sigma_\mathbf{x}+ \Sigma_\mathbf{y}}{2} \right|}^{-\frac{1}{2}}\int_{0}^{+\infty}\exp\left(-\frac{Q_{\mathbf{xy}}(\mathbf{x}-\mathbf{y})}{t^2}\right)\xi(dt)\\
&\propto& \pi^{-\frac{p}{2}}{\left| \frac{\Sigma_\mathbf{x}+ \Sigma_\mathbf{y}}{2} \right|}^{-\frac{1}{2}}R^S\left(\sqrt{Q_{\mathbf{xy}}(\mathbf{x}-\mathbf{y})}\right).
\end{eqnarray*}

Hence,
\begin{eqnarray*}\label{EqC13}
R^{NS}(\mathbf{x},\mathbf{y})=\frac{C^{NS}(\mathbf{x},\mathbf{y})}{\sqrt{C^{NS}(\mathbf{x},\mathbf{x})}\sqrt{C^{NS}(\mathbf{y},\mathbf{y})}}=\phi_{\mathbf{xy}}R^S\left(\sqrt{Q_{\mathbf{xy}}(\mathbf{x}-\mathbf{y})}\right). 
\end{eqnarray*}

For the proof of the corollaries \ref{Cor2} and \ref{Cor3}, let $\displaystyle{\tilde{C}^{NS}(\mathbf{x},\mathbf{y})=\int_{0}^{+\infty}g(\mathbf{x};t)g(\mathbf{y};t)t^{-p}\exp\left(-\frac{Q_{\mathbf{xy}}(\mathbf{x}-\mathbf{y})}{t^2}\right)M(dt)}$.

\subsubsection*{Proof of Corollary \ref{Cor2}}

If $g(\mathbf{x};t) \propto t^{\nu(\mathbf{x})}, \nu(\mathbf{x})>0, \forall \mathbf{x} \in \mathds{R}^p$ and $ M(dt)=2t^{p-1}h(t^2)\mathds{1}_{[0,+\infty)}(t)dt$, with $h(.)$ the density of the Gamma distribution $\mathcal{G}a(1,1/4a^2), a>0$, then
\begin{eqnarray*}\label{EqA14}
\tilde{C}^{NS}(\mathbf{x},\mathbf{y})=\int_{0}^{+\infty}\frac{1}{2a^2} t^{\nu(\mathbf{x})+\nu(\mathbf{y})-1}\exp\left(-\left(\frac{Q_{\mathbf{xy}}(\mathbf{x}-\mathbf{y})}{t^2}+\frac{t^2}{4a^2}\right)\right)dt.
\end{eqnarray*}

By the following change of variable, $w=\frac{t^2}{4a^2}$ and using an integral expression of the Bessel function \citep{Gra07,McL82}, we obtain
\begin{eqnarray*}\label{EqA15}
\tilde{C}^{NS}(\mathbf{x},\mathbf{y})
&=& \displaystyle{\int_{0}^{+\infty}(4a^2w)^{{(\nu(\mathbf{x})+\nu(\mathbf{y})-2)}/2}\exp\left(-w-\frac{Q_{\mathbf{xy}}(\mathbf{x}-\mathbf{y})}{4a^2w}\right)dw }\\
&=& \displaystyle{{(2a)}^{2(\nu(\mathbf{x},\mathbf{y})-1)}\int_{0}^{+\infty}w^{\nu(\mathbf{x},\mathbf{y})-1}\exp\left(-w-\frac{{(\sqrt{Q_{\mathbf{xy}}(\mathbf{x}-\mathbf{y})}/a)}^2}{4w}\right)dw} \\
&=& \displaystyle{2^{\nu(\mathbf{x},\mathbf{y})-1}a^{2(\nu(\mathbf{x},\mathbf{y})-1)}{\left(\frac{\sqrt{Q_{\mathbf{xy}}(\mathbf{x}-\mathbf{y})}}{a}\right)}^{\nu(\mathbf{x},\mathbf{y})}K_{\nu(\mathbf{x},\mathbf{y})}\left(\frac{\sqrt{Q_{\mathbf{xy}}(\mathbf{x}-\mathbf{y})}}{a}\right)}.
\end{eqnarray*}

We have
\begin{eqnarray*}\label{EqA16}
C^{NS}(\mathbf{x},\mathbf{y})
= \pi^{-\frac{p}{2}}{\left| \frac{\mathbf{\Sigma}_\mathbf{x}+ \mathbf{\Sigma}_\mathbf{y}}{2} \right|}^{-\frac{1}{2}}\tilde{C}^{NS}(\mathbf{x},\mathbf{y}).
\end{eqnarray*}

Hence,
\begin{eqnarray*}\label{EqA17}
R^{NS}(\mathbf{x},\mathbf{y})=\phi_{\mathbf{xy}}\frac{2^{1-\nu(\mathbf{x},\mathbf{y})}}{\sqrt{\Gamma(\nu(\mathbf{x}))\Gamma(\nu(\mathbf{y}))}}{\left(\frac{\sqrt{Q_{\mathbf{xy}}(\mathbf{x}-\mathbf{y})}}{a}\right)}^{\nu(\mathbf{x},\mathbf{y})}K_{\nu(\mathbf{x},\mathbf{y})}\left(\frac{\sqrt{Q_{\mathbf{xy}}(\mathbf{x}-\mathbf{y})}}{a}\right).
\end{eqnarray*}

\subsubsection*{Proof of Corollary \ref{Cor3}}

If $g(\mathbf{x};t) \propto t^{-\alpha(\mathbf{x})}, \alpha(\mathbf{x})>0, \forall \mathbf{x} \in \mathds{R}^p$ et $ M(dt)=2t^{p+3}h(t^2)\mathds{1}_{[0,+\infty)}(t)dt$, with $h(.)$ the density of the inverse Gamma distribution $\mathcal{IG} (1,a^2), a>0$, then
\begin{eqnarray*}\label{EqA18}
\tilde{C}^{NS}(\mathbf{x},\mathbf{y})
=\int_{0}^{+\infty}{2a^2} t^{-\alpha(\mathbf{x})-\alpha(\mathbf{y})-1}\exp\left(-\left(\frac{Q_{\mathbf{xy}}(\mathbf{x}-\mathbf{y})}{t^2}+\frac{a^2}{t^2}\right)\right)dt.
\end{eqnarray*}

By the following change of variable,$w=\frac{1}{t^2}$, we get
\begin{eqnarray*}\label{EqA19}
\tilde{C}^{NS}(\mathbf{x},\mathbf{y})
&=& a^2\int_{0}^{+\infty}w^{{(\alpha(\mathbf{x})+\alpha(\mathbf{y})-2)}/2}\exp\left(-\left(a^2w+Q_{\mathbf{xy}}(\mathbf{x}-\mathbf{y})w\right)\right)dw\\
&=& {a}^{-2(\alpha(\mathbf{x},\mathbf{y})-1)}\Gamma(\alpha(\mathbf{x},\mathbf{y}))\int_{0}^{+\infty}\exp\left(-Q_{\mathbf{xy}}(\mathbf{x}-\mathbf{y})w\right)\frac{a^{2\alpha(\mathbf{x},\mathbf{y})}}{\Gamma(\alpha(\mathbf{x},\mathbf{y}))}w^{\alpha(\mathbf{x},\mathbf{y})-1}\exp\left(a^2w\right)dw\\
&=& {a}^{-2(\alpha(\mathbf{x},\mathbf{y})-1)}\Gamma(\alpha(\mathbf{x},\mathbf{y}))\mathds{E}\left(\exp\left(-Q_{\mathbf{xy}}(\mathbf{x}-\mathbf{y})W\right) \right), \  W \sim \mathcal{G}a(\alpha(\mathbf{x},\mathbf{y}),a^2) \\
&=& {a}^{-2(\alpha(\mathbf{x},\mathbf{y})-1)}\Gamma(\alpha(\mathbf{x},\mathbf{y})){\left(1+\frac{Q_{\mathbf{xy}}(\mathbf{x}-\mathbf{y})}{a^2}\right)}^{-\alpha(\mathbf{x},\mathbf{y})}. 
\end{eqnarray*}

We have
\begin{eqnarray*}\label{EqA20}
C^{NS}(\mathbf{x},\mathbf{y})
= \pi^{-\frac{p}{2}}{\left| \frac{\mathbf{\Sigma}_\mathbf{x}+ \mathbf{\Sigma}_\mathbf{y}}{2} \right|}^{-\frac{1}{2}}\tilde{C}^{NS}(\mathbf{x},\mathbf{y}).
\end{eqnarray*}

Hence,
\begin{eqnarray*}\label{EqA21}
R^{NS}(\mathbf{x},\mathbf{y})=\phi_{\mathbf{xy}}\frac{\Gamma(\alpha(\mathbf{x},\mathbf{y}))}{\sqrt{\Gamma(\alpha(\mathbf{x}))\Gamma(\alpha(\mathbf{y}))}}{\left(1+\frac{Q_{\mathbf{xy}}(\mathbf{x}-\mathbf{y})}{a^2}\right)}^{-\alpha(\mathbf{x},\mathbf{y})}.
\end{eqnarray*}

\bigskip

\bibliographystyle{model1-num-names}

\end{document}